# Title: When heat goes astray - non-local heating in a semiconductor

**Authors:** M. Elhajhasan[1]*, E. Trukhan[3], K. Dudde[1], G. Würsch[1], J. Lierath[1], I. Rousseau[2], R. Butté[2], N. Grandjean[2], N. H. Protik[3], G. Romano[4], G. Callsen[1]*

**Affiliations:**

[1]Institut für Festkörperphysik, Universität Bremen; Bremen, 28359, Germany.

[2]Institute of Physics, École Polytechnique Fédérale de Lausanne (EPFL); Lausanne, 1015, Switzerland.

[3]Institut für Physik and Center for the Science of Materials Berlin (CSMB), Humboldt-Universität zu Berlin; Berlin, 12489, Germany.

[4]MIT-IBM Watson AI Lab, Massachusetts Institute of Technology (MIT); Cambridge, 02141, USA.

*Corresponding authors. Email: elhajmah@uni-bremen.de (M.E.); Email: gcallsen@uni-bremen.de (G.C.)

**Abstract:**

Heating of semiconductor devices limits their performance and lifetime, which must be addressed by thermal management starting at the heat source. It is a common assumption that the heat source and the resulting heat spot locally coincide, if their size exceeds the mean free paths of the main heat carriers, the phonons. We show that this paradigm of heat locality breaks down on length scales spanning several micrometers. As a consequence, non-local heating occurs in contradiction to Fourier's law. Therefore, we heat laterally structured semiconductor membranes that feature a rising number of interfaces with a well-focussed laser and map-out lattice temperatures by Raman thermometry. Remarkably, the non-local heating can exceed the laser-induced local heating, which we attribute to ballistic phonon transport far above cryogenic temperatures.

## Main Text:

The race toward ever smaller structures in the fields of photonics and electronics demands pathways for efficient heat dissipation (*1–3*). Poor thermal management of semiconductor micro- and nano-structures limits their performance (*4*, *5*) and lifetime (*6*, *7*). Device failure should be avoided by all means in light of an energy and resource-intensive fabrication, recycling, and disposal (*8*, *9*). Recent advances for the cooling of semiconductor microchips were, e.g., based on fluids in microchannels (*10*) or integrated thermoelectric coolers (*11*). However, ideally, thermal management should start at the level of the heat source, which is often of sub-micrometer size.

During the last decades, the challenging physics of thermal transport on the micro- and nano-scale (*12*) thwarted this aim. In most raw semiconductor materials, the transport of thermal energy is predominantly mediated by lattice vibrations called phonons, while the corresponding contribution of electrons often remains negligible, in contrast to the situation in metals (*13*). But modern semiconductor devices like transistors and laser diodes exhibit high doping concentrations and high current densities under operation, which ultimately requires the understanding of coupled electron and phonon propagation (*14*, *15*). Yet already the sole understanding of electron or phonon transport is challenging, because often only integrated quantities, such as, e.g., the electric resistivity $\rho$ or the thermal conductivity $\kappa$ are probed. The fundamental scattering processes that limit $\rho$ or $\kappa$, however, often remain obscured to the experimentalist.

To control transport properties, it is imperative to understand the various scattering mechanisms due to, e.g., defects, interfaces, and boundaries inherent to any real-world device. Insight into the intricate transport physics of electrons and phonons at the nanoscale can be gained by imaging their temperatures ($T_e$, $T_{ph}$) over a given spatial domain $(x,y)$, starting directly at the heat source. For the case of electron transport, it was shown that the temperature distribution of electrons $T_e(x,y)$ between closely spaced contacts can be mapped out by the scanning noise microscope (SNoiM) (*16*). Interestingly, the hottest spot in $T_e(x,y)$ was found 200 – 300 nm afar from the center of the constriction between the contacts, indicating a non-local heating phenomenon. Weng *et al.* interpreted this as a reminiscence of ballistic electron transport, whereby electrons travel unhindered until they deposit their thermal energy at a remote location, yielding electronic temperatures that can even exceed 2000 K (*16*). This observation can fundamentally improve the future design of electronic circuitry, which previously was always based on the assumption that the strongest self-heating occurs at the locations with the highest carrier density.

Close to room temperature ($T_{amb}$ = 295 K) or even at typical operation temperatures of high power electronics (*17*), micro-LEDs (*18*), or nanolasers (*19*), it remains an open question, whether such non-local heating phenomena also appear for phonons. In the classic Fourier picture of purely diffusive phonon transport, the temperature at a single heat source $T_s$ always exhibits the highest value and the local heat flux is proportional to the temperature gradient and $\kappa$. In this picture, a defect, an interface, an inhomogeneity of the material, or a boundary can disturb the flow of heat, but $T_s$ remains the highest temperature and with increasing distance to a single heat source one can only observe a monotonic temperature decrease. In contrast, in this work, we show that this statement does not hold as soon as the injected energy is partially channeled into phonons that propagate ballistically, leading to non-local heating in a semiconductor even at $T_{ph} \gg 295$ K, clearly violating the aforementioned monotony of the temperature evolution and therefore Fourier's law.

Three key ingredients are needed to reveal non-local heating that goes beyond Fourier's law. First, $T_{ph}(x,y)$ needs to be imaged around a local heat source, which we achieve by our potent thermal imaging system (*20*) based on two-laser Raman thermometry (2LRT) (*21–23*). For 2LRT a heat laser is focused onto the sample (heat laser wavelength $\lambda_{heat}$ = 266 nm), while a second probe laser (probe laser wavelength $\lambda_{probe}$ = 488 nm) is scanned around the laser-induced heat spot to measure $T_{ph}(x,y)$ via the Raman signal as illustrated in Fig. 1A. Analyzing semiconductor membranes is beneficial, because the thermal transport mostly occurs in-plane, which is in line with our $T_{ph}(x,y)$ mapping capabilities. Second, one requires a semiconductor material in which phonons with sufficiently long mean free paths (MFP) contribute significantly to thermal transport at $T_{ph} \geq 295$ K for the given optical heating. For this study, we chose wurtzite GaN, a well-established material that is widely used in photonics and electronics due to its excellent optical, electrical, and thermal properties (*24*). In an earlier work, we already demonstrated the relevance of large MFP values in ≈ 250-nm-thick membranes primarily made from wurtzite GaN (*20*), which also form the basis for this study. Third, a thermal resistance like, e.g., a crystal defect or a boundary needs to be introduced next to the laser-induced heat source. We introduce lateral boundaries (counted by $n$) in our membranes in close distance ($d \approx 2 - 5$ µm) to the heat laser spot, yielding a suspended edge ($n$ = 1), a corner ($n$ = 2), and a hexagon ($n$ = 6) as illustrated in Fig. 1A (bottom). With increasing $n$, we strengthen the effect of non-local heating in our structures, while for n = 6 and $d \approx 2$ µm, even $T_{ph}(edge) >> T_s$ can be observed. Motivated by our experiments, we call this phenomenon "edge heating" and interpret it as a consequence of ballistic phonon transport that is promoted by our optical heating up to ~ 970 K, before device failure occurs.

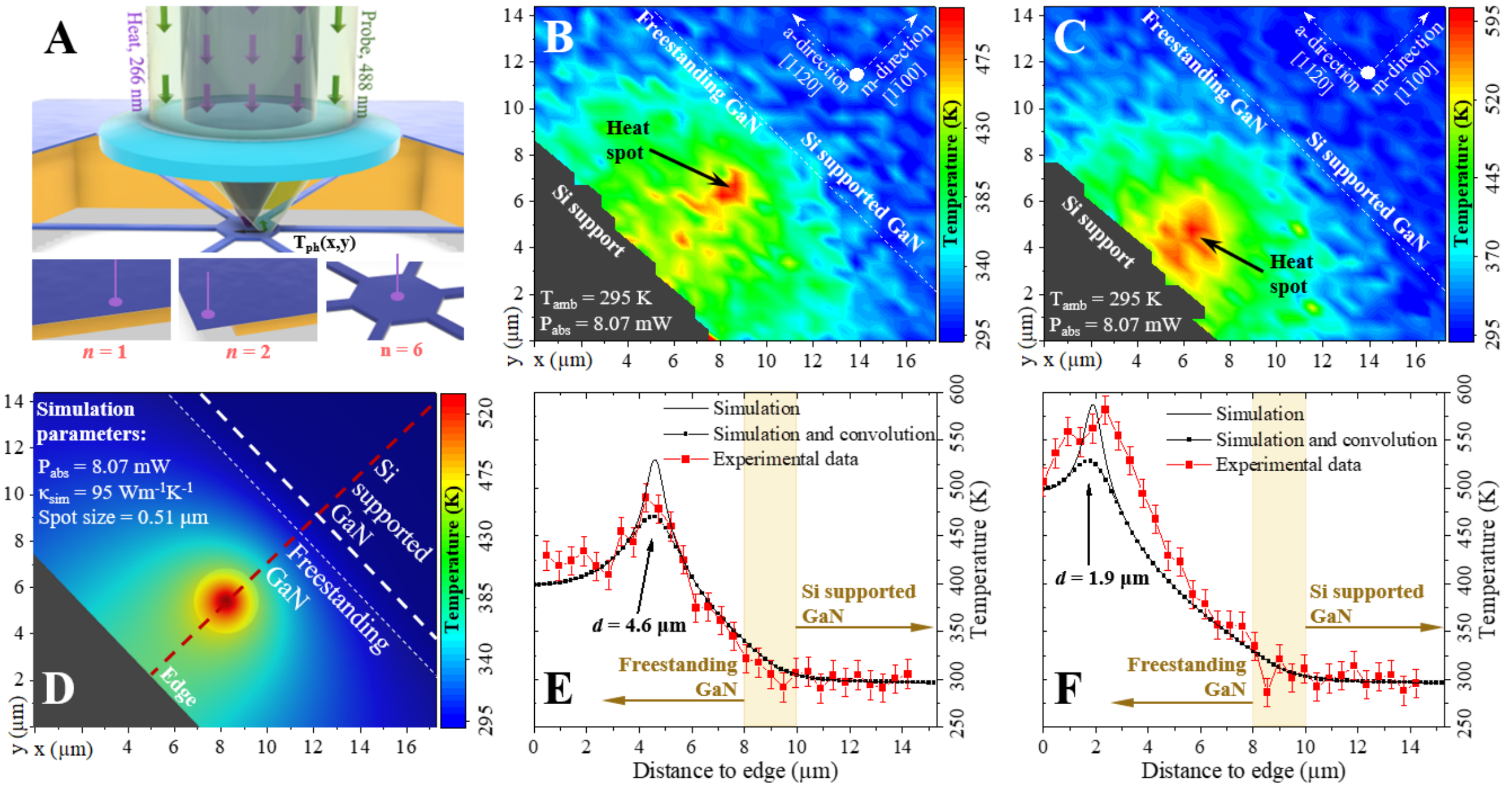

**Fig. 1. Thermal imaging of a membrane edge by two-laser Raman thermometry (2LRT).**
(**A**) Sketch of the 2LRT setup and schemes of the suspended structures analyzed in this study. A growing number of boundaries $n$ was processed into a membrane mostly made of GaN, yielding an edge ($n$ = 1), a corner ($n$ = 2), and a hexagon ($n$ = 6). The setup and sample details can be found in fig. S1 and section II of (*20*). (**B**) Temperature mapscan $T_{ph}(x,y)$ for a fixed distance $d$ = 4.6 ± 0.3 µm between the heat spot and the edge of the membrane. The absorbed power of the heat laser is given by $P_{abs}$. The sample has been designed so that the sides of the structure coincide with the in-plane crystal directions (see top right of the image). (**C**) $T_{ph}(x,y)$ for $d$ = 1.9 ± 0.3 µm. (**D**) 3D simulation of the experiment (top-view) based on the heat equation implemented in COMSOL. The thermal conductivity $\kappa_{sim}$ was measured in (*20*), which also describes the determination of the absorbed power $P_{abs}$ of the heat laser. The red dashed line indicates the location of the cutline that is compared to the experiment. (**E, F**) Comparison of the experimental and theoretical temperature cutlines through the heat laser spot perpendicular to the membrane edge for $d$ = 1.9 ± 0.3 µm (**E**) and $d$ = 4.6 ± 0.3 µm (**F**). The shaded beige rectangles designate the transition area between the supported and freestanding part of the membrane (fig. S2). It corresponds to the area between the thick and thin dashed white lines in **D**. All experimental data sets were recorded in vacuum (< 3 × $10^{-6}$ mbar) at $T_{amb}$ = 295 K.

## Imaging temperature distributions with sub-µm resolution

The first indication for the edge heating phenomenon is given by the temperature mapscans $T_{ph}(x,y)$ shown in Fig. 1B and C. These maps were recorded for two distances $d$ between the heat spot and the edge ($n$ = 1) of the membrane ($d$ = 4.6 ± 0.3 µm and 1.9 ± 0.3 µm) at $T_{amb}$ = 295 K, while the sample was placed in vacuum (< 3 × $10^{-6}$ mbar) for all measurements to avoid convection cooling. $T_{ph}$ can be extracted from both, the Raman mode position $\omega(T_{ph})$ and the width of the mode $\Delta\omega(T_{ph})$. A temperature calibration of these quantities was obtained from measurements on the freestanding part of our membrane in a heat stage (fig. S3). For Fig. 1B and C we extracted $T_{ph}$ based on $\omega$. The long recording times for a single $T_{ph}(x,y)$ map (≈ 2 h to 10 h) require sub-micron position stability of the heat laser, the probe laser, and the sample. This challenge motivated the construction of a specifically designed 2LRT setup. Relevant details about the sample, the 2LRT measurements, and the simulations can be found in the Supplementary Materials (SM) and in (*20*). We first simulated our experiments using the 3D heat equation implemented in COMSOL (corresponds to the "Fourier picture") as exemplified in Fig. 1D for $d$ = 4.6 ± 0.3 µm (fig. S4A shows the result for $d$ = 1.9 ± 0.3 µm). The thermal conductivity of the membrane ($\kappa_{sim}$) and the absorbed laser power ($P_{abs}$) are both measured input parameters for the simulations, cf. sections V.A.1 and V.A.2 of (*20*). A comparison between the experimental (red datapoints) and numerical results (solid black lines), which are based on a cutline through the heat spot (red dashed line in Fig. 1D), is shown in Fig. 1E ($d$ = 4.6 ± 0.3 µm) and 1F ($d$ = 1.9 ± 0.3 µm). For $d$ = 4.6 ± 0.3 µm, we can achieve a good agreement between experiment and theory, as soon as the experimental averaging induced by the measured finite size of the focused probe laser spot diameter is considered (connected black datapoints, see SM and (*20*) for more details). In Fig. 1E, only a subtle discrepancy between experiment and theory remains at the edge of the membrane. This discrepancy worsens when reducing $d$, as shown in Fig. 1F. There, the experimental $T_{ph}$ distribution is significantly broader than the theoretical prediction and its maximal value exceeds its theoretical counterpart around the heat spot. The presence of an edge appears as a disturbance to the heat flow in a way that is not covered by our modelling based on the Fourier equation.

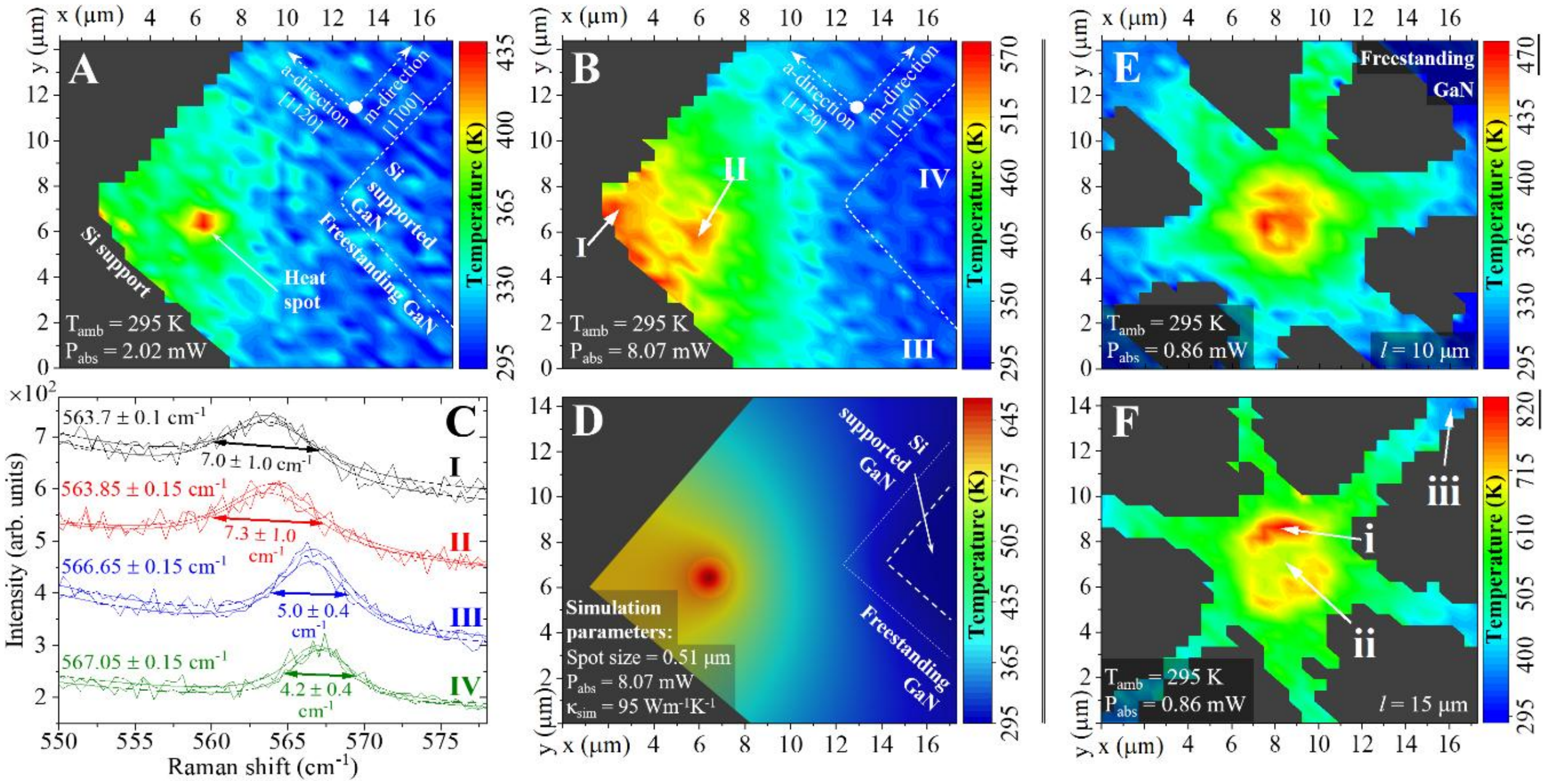

**Fig. 2. Thermal imaging of a membrane corner and two suspended hexagons by 2LRT.**
(**A**) Temperature mapscan at a corner ($n$ = 2) with the heat laser placed at $d$ = 3.5 ± 0.3 µm from both edges and $P_{abs}$ = 2.02 mW. (**B**) Increasing $P_{abs}$ to the same value as used in Fig. 1 (8.07 mW) yields a $T_{ph}(x,y)$ map that shows heating at the position of the heat laser (II) and at the edge of the membrane (I) in the corner region, where $T_{ph}(edge) \approx T_s$. (**C**) Raman spectra showing the optical $E_2^{high}$ mode of GaN in hot (I and II) and cold (III and IV) positions of the membrane that are indicated in **B**. Two spectra (from closely matching locations) and two fits to the Raman mode are overlaid for each position I – IV. (**D**) Corresponding simulation based on the 3D heat equation in COMSOL. No additional heating at the edges is visible here. (**E**) Temperature mapscan of a suspended hexagon ($n$ = 6, compare with Fig. 1A) with the heat laser spot positioned at its center (distance to the perimeter of the hexagon $d \approx 2.0 \pm 0.3$ µm) and $P_{abs}$ = 0.86 mW. Six nanobeams with an arm length $l$ = 10 µm hold the hexagon. (**F**) While keeping $P_{abs}$ constant, $l$ can be increased to reach higher temperatures. As a result, an indication for the edge heating appears for $l$ = 15 µm. For a support temperature of 295 K, neither $P_{abs}$ nor $l$ can be further increased without damaging the structure before an even clearer edge heating occurs (i - iii indicate the recording positions for the corresponding Raman spectra shown in fig. S6 B).

## Rising the number of boundaries

To further investigate this anomalous heating at the edge in a systematic way, we measured the processed corners ($n$ = 2) of our membrane and imaged $T_{ph}(x,y)$ for constant $d$ = 3.5 µm ± 0.3 µm value with respect to both edges, and varied $P_{abs}$ values as shown in Fig. 2A and B (intermediate $P_{abs}$ step shown in fig. S5B). Edge heating cannot be observed in Fig. 2A for $P_{abs}$ = 2.02 mW. Repeating the measurement with $P_{abs}$ = 8.07 mW yields a different story. The temperature at the edge reaches the temperature at the heat laser spot, which is inconceivable in a purely diffusive, Fourier-like, transport regime. On this few-µm-scale, which is on the order of typical MFP values in GaN (*20*), a sufficiently large fraction of the energy deposited by the heat laser must have been carried by ballistically traveling phonons to the edge of the sample. Here these phonons scatter without having dissipated much momentum or energy along the way. To solidify our observation, we show Raman spectra in Fig. 2C for selected hot (I and II) and cold (III and IV) positions, which are indicated in Fig. 2B. As a sign of heating, the non-polar $E_2^{high}$ Raman mode of GaN is clearly shifted to lower relative wavenumbers ($\omega$) at positions I and II, in contrast to the colder positions III and IV. This temperature induced Raman mode shift $\omega(T_{ph})$ is caused by the net volumetric thermal expansion of GaN upon heating in combination with 3-, 4-, and higher order phonon scattering corrections to the mode self-energy (*25*). Simultaneously, the Raman mode broadens at the hotter positions, which we quantify via the corresponding full width at half maximum (FWHM) $\Delta\omega$ of the peak. The optical $E_2^{high}$ phonon of GaN that serves as a temperature probe for our 2LRT maps increasingly scatters with other phonons as soon as the phonon bath gets more populated. This results in the reduction of the lifetime of the mode, which translates into an increase of $\Delta\omega$ (see Fig. 2C for positions I and II) via the time-energy uncertainty principle. In a first order approximation, $\Delta\omega$ is not affected by any volumetric effects (*26*, *27*). The corresponding $\Delta\omega(x,y)$ map can also be converted into a temperature map $T_{ph}(x,y)$ that again shows edge heating (fig. S5A). Furthermore, the onset of the edge heating for $n = 2$ is already visible in fig. S5B for an intermediate $P_{abs}$ value of 4.03 mW. Based on the Raman mode shift, any impact of built-in strain on $\omega$ has been removed from our $T_{ph}(x,y)$ data by subtracting a heated from an unheated $\omega(x,y)$ map before any conversion to $T_{ph}(x,y)$ was performed [see section VI of (*20*)]. In addition to the built-in strain, temperature-induced strain can still impact Raman thermometry based on $\omega$ (*28*). Thermal expansion leads to the built-up of compressive strain in the position of the heat laser, causing a shift of the $E_2^{high}$ mode to larger relative wavenumbers (*26*, *27*), which partially compensates the effect of temperature on $\omega$. The resulting conversion of $\omega$ to temperatures leads to a lowering of the measured temperatures in the laser spot [sections V. A. 1 and V. A. 2 of (*20*) and section 4 of (*28*)]. Previously, we showed in (*28*) that this is a few percent effect in silicon, which therefore cannot explain the edge heating phenomenon. In addition, if the anomalous heating at the edge would be related to thermally induced strain, then it should not cause an enlargement of $\Delta\omega(x,y)$, which is, however, visible in Fig. 2C and in the corresponding temperature map shown in fig. S5A. Finally, Fig. 2B validates our previous observation from Fig. 1 that edges disturb the flow of heat, which leads to non-local heating.

## Edge heating as a signature of ballistic phonon transport

The most extreme example of the edge heating phenomenon is given by our $T_{ph}(x,y)$ imaging of ≈ 4-µm-wide hexagons ($n = 6$ , scanning electron microscope and light microscope images shown in fig. S1C - G), which are held by six nanobeams of varying length $l$ and fixed width (500 nm). First, for our shortest nanobeams ($l$ = 10 µm) and $P_{abs}$ = 0.86 mW, we obtain a 2LRT map (Fig. 2E), which does not show any pronounced edge heating (also valid for a $T_{ph}(x,y)$ map shown in fig. S7 for $l$ = 10 µm and $P_{abs}$ = 1.20 mW). By increasing $l$ from 10 µm to 15 µm while maintaining $P_{abs}$ = 0.86 mW, the maximum temperature of the structure rises from ≈ 470 K to 820 K and the edge heating reappears as shown in Fig. 2F. Here, in contrast to the edge ($n = 1$) and the corner ($n = 2$) configurations, the boundaries of the hexagon are positioned in the direction of the heat flux toward the heat sink provided by the Si substrate. This experimental configuration of the hexagons ($n = 6$) seems particularly well suited to reveal the edge heating phenomenon. Similar to Fig. 2B, local Raman spectra acquired at the heated edge (i), the center of the hexagon (ii), and a position on a nanobeam (iii) confirm the $T_{ph}(x,y)$ distribution shown in Fig. 2F based on $\omega$ (fig. S6B shows the corresponding local Raman spectra i – iii).

Recording $T_{ph}(x,y)$ maps for larger $l$ or $P_{abs}$ values proved experimentally challenging as long as the temperature of the supporting Si substrate remained at $T_{amb}$ = 295 K. The experimental conditions associated to Fig. 2F approach the destruction limit of the structure at the given long recording times, which troubled the acquisition of highly spatially resolved $T_{ph}(x,y)$ maps. To circumvent this issue, we cooled the Si substrate to ≈ 12 K, which enabled the recording of Fig. 3A for $l$ = 30 µm and $P_{abs}$ = 0.86 mW. Here, we obtain a more symmetric edge heating that seems to form a structured ring. Note that despite the cooling, the imaged section of the hexagon with its nanobeams still shows heating in the range of ≈ 450 K – 880 K, exceeding the temperatures shown in Fig. 2F due to the enlargement of $l$ from 15 µm to 30 µm. The positions of the hexagon's boundaries have been indicated in Fig. 3A by white dashed lines. These boundary positions were extracted from maps of the relative reflectivity changes ($\Delta R(x,y)/R$) of the Raman laser when scanning over the structure, which are always acquired for each 2LRT map. The ring-like edge heating follows the contour of the structure. This observation is supported by a corresponding $T_{ph}(x,y)$ map in fig. S6A based on $\Delta\omega$, which also shows the heated contour.

To gain further insights into the edge heating of the hexagon, we recorded a $T_{ph}(x,y)$ map based on $\omega$ with increased spatial resolution as shown in Fig. 3B, where the average step size was reduced from 530 nm to 265 nm. Based on Fig. 3A and B we can identify three key experimental findings, which are: **F1** - the edge heating itself (e.g., defined by a non-monotonous temperature evolution with increasing distance to our *single*, laser-induced heat source); **F2** - $T_{ph}(edge) >> T_s$; and **F3** - the pronounced temperature drop at the nanobeam interconnects. Any modeling of the edge heating phenomenon for the hexagons must at least qualitatively describe these key experimental findings **F1** – **F3**. Note that the maximum value for $T_{ph}(edge)$ is higher in Fig. 3B (≈ 970 K) compared to Fig. 3A (≈ 870 K), because of the increased spatial resolution.

In a previous work (*20*), we already excluded any significant impact of non-phononic energy transport phenomena in our optically heated semiconductor membranes. Transport contributions of photons, electrons, and excitons are negligible, as summarized in S-Sec. II of (*20*). At the current stage, phonon-polariton (pp) transport cannot be completely excluded, although this type of transport commonly yields thermal conductivities in the range of $\kappa_{pp} \approx 1 – 5$ Wm$^{-1}$K$^{-1}$ (*29*, *30*) for $SiO_2$ thin films, which is smaller than any phonon-induced heat conduction in our III-nitride membrane, where we found $\kappa_0 \approx 95$ $Wm^{-1}K^{-1}$ (*20*).

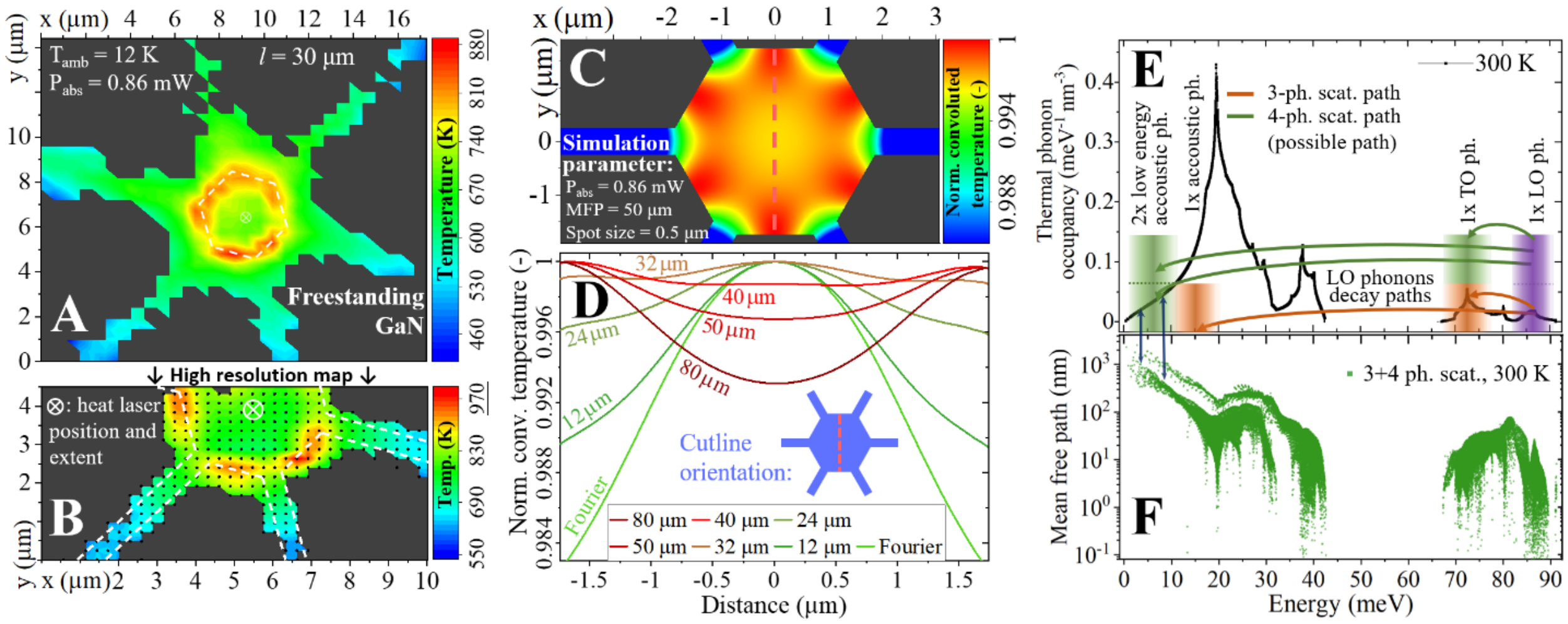

**Fig. 3. 2LRT imaging of suspended GaN hexagons, 2D-BTE simulations, and scheme of the LO phonon decay.** (**A**) Temperature mapscan of a suspended hexagon ($n$ = 6) that shows the edge heating for $T_{amb}$ = 12 K (Si substrate temperature) and $l$ = 30 µm. All other experimental parameters are the same as for Fig. 2E and F. Note that despite of the substrate cooling, the suspended hexagon's temperatures exceed the ones shown in Fig. 2F. See the main text for details. (**B**) High resolution $T_{ph}(x,y)$ scan of the lower half of the hexagon shown in **A**. Black dots indicate the location of the data points and the structure's contour is approximated by the white dashed lines. The edge heating appears close to the perimeter of the hexagon and weakens at the nanobeam interconnects. (**C**) Convoluted simulated temperature map based on the single MFP 2D-Boltzmann transport equation (2D-BTE) for the hexagon and its nanobeam support. Despite the simplicity of this model it reproduces three key experimental findings (edge heating, $T_{ph(edge)} >> T_s$, temperature drop at the nanobeam interconnects). The red dashed line indicates the location of the data displayed in (**D**), where 2D-BTE simulations are compared to a Fourier simulation. The simulated temperature has been normalized and convoluted with a 2D Gaussian to account for the probe laser's focus spot size. (**E**) Thermal phonon occupancy per unit cell for GaN at $T$ = 300 K. The colored columns and arrows indicate some possible decay paths of LO phonons via 3-phonon (orange) and 4-phonon (green) scattering. (**F**) MFP values vs. phonon energy in wurtzite, bulk GaN.

We followed several pathways with increasing level of complexity to reproduce our measured $T_{ph}(x,y)$ maps through numerical simulations assuming phonon-induced heat conduction. First, we used a simple Fourier model with a constant thermal conductivity $\kappa_0$, which naturally did not capture the edge heating. Second, we introduced a temperature dependent $\kappa(T)$ into this model to account for the non-linear behavior of the heat equation in the presence of strong temperature gradients (*31*). Also, these simulations did not reproduce the experimental temperature profiles, yet the maximal simulated temperature rise approached the experimental findings. A summary of our linear and non-linear Fourier modeling is given in fig. S8.

Finally, we turned to the 2D-Boltzmann transport equation (2D-BTE) to model the thermal transport in our suspended GaN hexagons ($n = 6$) from Fig. 3. We used the open source package OpenBTE to solve the BTE (*32*, *33*). Given the complex geometry of our sample, we performed simulations with one phonon mode at a time with MFP values scaling from 8 – 80 µm (see Fig. 3D and fig. S9 for all MFP increments) in addition to a Fourier simulation, which was used to seed our BTE solver. In agreement with our experiments, the heat spot diameter is set to 0.5 µm ($P_{abs}$ = 0.86 mW), while all sample boundaries were approximated as being diffusive. More details regarding the theoretical framework of our 2D-BTE simulations can be found in the SM. In Fig. 3C we show a theoretical $T_{ph}(x,y)$ map that originates from a convolution of a 2D-BTE simulation output for MFP = 50 µm and a 2D Gaussian that accounts for the experimental beam profile of the focused probe laser. Despite the simplified physics in our challenging simulations resulting in Fig. 3C, we can qualitatively reproduce our three key experimental findings **F1** – **F3**, which supports our interpretation of the edge heating.

Extracting normalized, convoluted temperature cutlines from our simulations at the location shown by the red dashed line in Fig. 3C, enables the comparison of our calculated results for a larger set of MFP values shown in Fig. 3D. Here, we can see that the edge heating gradually appears as the MFP is increased and $T_{ph}(edge) >> T_s$ (**F2**) is reached for MFP values > 32 µm. From an experimental point of view, one would expect that phonons with MFP values close to the distance between the heat source and the heated boundary (e.g., MFP ≈ 2 µm, equaling half of the diameter of the hexagon) suffice to cause edge heating. We point out, however, that several simplifications in our 2D-BTE model prevent quantitative agreement with our experimental observations. First, as our simulations are 2D, the effect of thickness is not modeled, potentially neglecting further insights on the MFP threshold triggering the edge heating. Second, phonon generation upon optical heating is not considered. Third, the use of a single-MFP model is also a major limitation, as it does not capture the interplay between phonons with different MFPs and the boundary (*34*). Lastly, the employed 2D-BTE model is linear, while the wide range of temperatures observed experimentally suggests the onset of a non-linear regime, where material properties become temperature-dependent. Interestingly, edge heating was theoretically predicted in at least two purely theoretical studies on porous materials (Fig. 5 of (*12*) and Fig. 2C of (*35*)). Additional theoretical results and a detailed explanation of our 2D-BTE modeling is given in the SM.

### High temperature phenomenon

The recorded $T_{ph}(x,y)$ maps indicate the presence of phonons with high MFP values in our membranes even at temperatures above 295 K. This phenomenon is connected to our optical heating and the appearance of higher order phonon-phonon scattering events (4-phonon and above) at elevated temperatures. For $T_{ph}(x,y)$ maps displaying lower maximal temperatures (< 500 K), no clear edge heating is visible (e.g. Fig. 2E and fig. S7). The phenomenon becomes prominent at temperatures > 500 K (e.g., Fig. 2B and F), independently of the laser power.

To shine more light on this particular observation, we plot the thermal phonon occupancy for 300 K over phonon energy for wurtzite GaN in Fig. 3E. Figure 3F additionally shows the corresponding energy-resolved MFP distribution of phonons at 300 K. Details of the underlying *ab initio* theory are given in the SM. The optical excitation of our sample causes excited electrons to release the absorbed energy to the lattice mostly by creating LO phonons (Fig. 7 of (*40*)). Fig. 3E (see purple column) in connection with Fig. 3F shows that LO phonons in wurtzite GaN do not exhibit sufficient MFP values to cause the edge heating (LO MFP values ≈ 1 – 30 nm). In a low heating situation (< 500 K), where 3-phonon scattering is dominant, the conservation of energy and momentum imposes strict decay paths for LO phonons, which are summarized in table S1 and sketched (orange arrow) in Fig. 3E (*41*). Both transverse and longitudinal acoustic (TA and LA) phonons as well as phonons with the $E_2^{low}$ symmetry are generated by 3-phonon scattering, exhibiting maximal MFP values of 500 nm (see Fig. 3F), while most phonons exhibit MFP values between 10 to 100 nm. Such MFP values are not sufficient to explain the edge heating. Naturally, 3-phonon scattering processes contribute to the overall heating of our structure from Fig. 3. However, this conventional heating is overlaid by the edge heating phenomenon, which even surpasses $T_s$.

Thus, the edge heating must be linked to our optical heating and the appearance of higher order phonon-phonon scattering events (4-phonon and above) at elevated temperatures (> 500 K). At these temperatures, numerical simulations of $\kappa$ including 3- & 4-phonon scattering start to deviate from the experimental results (Fig. 4 of (*42*)), which we interpret as the onset of higher order scattering channels for LO phonons in wurtzite GaN. An exemplary 4-phonon-scattering process is illustrated by the green arrows in Fig. 3E. Only through such higher order scattering events one can reach acoustic phonons with a sufficiently large MFP values (0.5 – 3 μm) via the decay of LO phonons. As an example, the reachable phonon MFP interval is depicted by the deep-blue, vertical arrows that connect Fig. 3E and F (300 K). Similar data for 600 K (i.e., close to $T_s$ in Fig. 3A and B) shows that the maximal reachable MFP values for acoustic phonons is reduced to ≈ 605 nm. Thus, we cannot exclude that a limited number of 3- and 4-phonon scattering processes leads to the edge heating phenomenon, which would be supported by the heating at the position of the heat laser that is observed in, e.g., Fig. 2B as well as Fig. 3A and B. This explains why edge heating appears at elevated temperatures, regardless of the laser power. Additional decays of the transverse optical (TO) and $E_2^{high}$ phonons are not considered here. A microscopic, real-space description of the optically induced heat generation and the subsequent phonon transport via higher order scattering process in our structures goes beyond the present state of theory.

## Summary and conclusions

We developed a thermal imaging system to directly map phonon-induced non-local heating phenomena in a III-nitride semiconductor membrane. In addition to the common laser-induced heat spot, we observe a pronounced heating of boundaries (semiconductor/vacuum interfaces) in our membranes at a distance of a few micrometers. Starting at a temperature of about 500 K, we can observe phonons with large MFP values that travel ballistically to the edges of our membranes, where they scatter and deposit their energy. This gives rise to the edge heating phenomenon. Interestingly, this observation occurs significantly above room temperature and the edge heating can reach temperatures up to 970 K in our structures. We link this observation to the rising importance of 4- and higher order phonon scattering at such elevated temperatures, which enables the generation of low energy phonons with large MFP values. This observation clearly shows the importance of the particular heat generation process for any modeling of thermal transport. Overall, similar edges appear in any real-world device and must be considered for an effective thermal management. Our observation of non-local heating suggests that device failure can appear a few micrometers away from the heat source, which is at the same time ideal to position a heat sink or spreader. Understanding this phenomenon opens the door for smart heat management approaches that directly address the heat source and all edges and interfaces in close proximity.

**References and Notes:**

1. A. L. Moore, L. Shi, Emerging challenges and materials for thermal management of electronics. *Materials Today* **17**, 163–174 (2014).

2. M. M. Waldrop, More than Moore. *Nature* **530**, 144–147 (2016).

3. X. Wu, L. Tang, C. L. Hardin, C. Dames, Y. Kodera, J. E. Garay, Thermal conductivity and management in laser gain materials: A nano/microstructural perspective. *Journal of Applied Physics* **131**B, 020902 (2022).

4. S. K. Mohanty, Y.-Y. Chen, P.-H. Yeh, R.-H. Horng, Thermal Management of GaN-on-Si High Electron Mobility Transistor by Copper Filled Micro-Trench Structure. *Sci Rep* **9**, 19691 (2019).

5. Y. P. Pundir, A. Bisht, R. Saha, P. K. Pal, Effect of Temperature on Performance of 5-nm Node Silicon Nanosheet Transistors for Analog Applications. *Silicon* **14**, 10581–10589 (2022).

6. P. K. Schelling, L. Shi, K. E. Goodson, Managing heat for electronics. *Materials Today* **8**, 30–35 (2005).

7. M. Meneghini, C. De Santi, I. Abid, M. Buffolo, M. Cioni, R. A. Khadar, L. Nela, N. Zagni, A. Chini, F. Medjdoub, G. Meneghesso, G. Verzellesi, E. Zanoni, E. Matioli, GaN-based power devices: Physics, reliability, and perspectives. *Journal of Applied Physics* **130**, 181101 (2021).

8. E. D. Williams, R. U. Ayres, M. Heller, The 1.7 Kilogram Microchip: Energy and Material Use in the Production of Semiconductor Devices. *Environ. Sci. Technol.* **36**, 5504–5510 (2002).

9. C. L. Gan, M.-H. Chung, Y.-S. Zou, C.-Y. Huang, H. Takiar, Technological sustainable materials and enabling in semiconductor memory industry: A review. *e-Prime - Advances in Electrical Engineering, Electronics and Energy* **5**, 100245 (2023).

10. R. Van Erp, R. Soleimanzadeh, L. Nela, G. Kampitsis, E. Matioli, Co-designing electronics with microfluidics for more sustainable cooling. *Nature* **585**, 211–216 (2020).

11. W. Fan, Z. Wu, S. Hong, K. Sheng, High-performance integrated thermoelectric coolers for electronics cooling. *Commun Mater* **6**, 114 (2025).

12. M.-S. Jeng, R. Yang, D. Song, G. Chen, Modeling the Thermal Conductivity and Phonon Transport in Nanoparticle Composites Using Monte Carlo Simulation. *Journal of Heat Transfer* **130**, 042410 (2008).

13. D. G. Cahill, K. Goodson, A. Majumdar, Thermometry and Thermal Transport in Micro/Nanoscale Solid-State Devices and Structures. *Journal of Heat Transfer* **124**, 223–241 (2002).

14. N. H. Protik, B. Kozinsky, Electron-phonon drag enhancement of transport properties from a fully coupled *ab initio* Boltzmann formalism. *Phys. Rev. B* **102**, 245202 (2020).

15. Y. Quan, B. Liao, Coupled Electron-Phonon Hydrodynamics in Two-Dimensional Semiconductors. *Phys. Rev. Lett.* **134**, 226301 (2025).

16. Q. Weng, S. Komiyama, L. Yang, Z. An, P. Chen, S.-A. Biehs, Y. Kajihara, W. Lu, Imaging of nonlocal hot-electron energy dissipation via shot noise. *Science* **360**, 775–778 (2018).

17. C. Buttay, D. Planson, B. Allard, D. Bergogne, P. Bevilacqua, C. Joubert, M. Lazar, C. Martin, H. Morel, D. Tournier, C. Raynaud, State of the art of high temperature power electronics. *Materials Science and Engineering: B* **176**, 283–288 (2011).

18. Y. Feng, M. Zhanghu, B.-R. Hyun, Z. Liu, Thermal characteristics of InGaN-based green micro-LEDs. *AIP Advances* **11,** 045227 (2021).

19. Q. Zhang, G. Li, X. Liu, F. Qian, Y. Li, T. C. Sum, C. M. Lieber, Q. Xiong, A room temperature low-threshold ultraviolet plasmonic nanolaser. *Nat Commun* **5,** 4953 (2014).

20. M. Elhajhasan, W. Seemann, K. Dudde, D. Vaske, G. Callsen, I. Rousseau, T. F. K. Weatherley, J.-F. Carlin, R. Butté, N. Grandjean, N. H. Protik, G. Romano, Optical and thermal characterization of a group-III nitride semiconductor membrane by microphotoluminescence spectroscopy and Raman thermometry. *Physical Review B* **108**, 235313 (2023).

21. M. Soini, I. Zardo, E. Uccelli, S. Funk, G. Koblmüller, A. Fontcuberta I Morral, G. Abstreiter, Thermal conductivity of GaAs nanowires studied by micro-Raman spectroscopy combined with laser heating. *Applied Physics Letters* **97**, 263107 (2010).

22. B. Stoib, S. Filser, J. Stötzel, A. Greppmair, N. Petermann, H. Wiggers, G. Schierning, M. Stutzmann, M. S. Brandt, Spatially Resolved Determination of Thermal Conductivity by Raman Spectroscopy. *Semicond. Sci. Technol.* **29**, 124005 (2014).

23. J. S. Reparaz, E. Chavez-Angel, M. R. Wagner, B. Graczykowski, J. Gomis-Bresco, F. Alzina, C. M. Sotomayor Torres, A novel contactless technique for thermal field mapping and thermal conductivity determination: Two-Laser Raman Thermometry. *Review of Scientific Instruments* **85**, 034901 (2014).

24. H. Morkoç, *Handbook of Nitride Semiconductors and Devices* (Wiley, Weinheim, 2009).

25. W. S. Li, Z. X. Shen, Z. C. Feng, S. J. Chua, Temperature dependence of Raman scattering in hexagonal gallium nitride films. *Journal of Applied Physics* **87**, 3332–3337 (2000).

26. T. Beechem, A. Christensen, S. Graham, D. Green, Micro-Raman thermometry in the presence of complex stresses in GaN devices. *Journal Of Applied Physics* **103**, 124501 (2008).

27. G. Callsen, J. S. Reparaz, M. R. Wagner, R. Kirste, C. Nenstiel, A. Hoffmann, M. R. Phillips, Phonon deformation potentials in wurtzite GaN and ZnO determined by uniaxial pressure dependent Raman measurements. *Applied Physics Letters* **98**, 061906 (2011).

28. K. Dudde, M. Elhajhasan, G. Würsch, J. Themann, J. Lierath, D. Paul, N. H. Protik, G. Romano, G. Callsen, Phonon mean free path spectroscopy by Raman thermometry. *Materials Today Physics* **57**, 101784 (2025).

29. L. Tranchant, S. Hamamura, J. Ordonez-Miranda, T. Yabuki, A. Vega-Flick, F. Cervantes-Alvarez, J. J. Alvarado-Gil, S. Volz, K. Miyazaki, Two-Dimensional Phonon Polariton Heat Transport. *Nano Lett.* **19**, 6924–6930 (2019).

30. D. Li, Z. Pan, J. D. Caldwell, Phonon polariton-mediated heat conduction: Perspectives from recent progress. *Journal of Materials Research* **39**, 3193–3201 (2024).

31. Q. Zheng, C. Li, A. Rai, J. H. Leach, D. A. Broido, D. G. Cahill, Thermal conductivity of GaN, GaN 71 , and SiC from 150 K to 850 K. *Phys. Rev. Materials* **3**, 014601 (2019).

32. G. Romano, OpenBTE: a Solver for ab-initio Phonon Transport in Multidimensional Structures. arXiv arXiv:2106.02764 [Preprint] (2021). https://doi.org/10.48550/arXiv.2106.02764.

33. G. Romano, S. G. Johnson, Inverse design in nanoscale heat transport via interpolating interfacial phonon transmission. *Struct Multidisc Optim* **65**, 297 (2022).

34. G. Romano, A. M. Kolpak, Diffusive Phonons in Nongray Nanostructures. *Journal of Heat Transfer* **141**, 012401 (2019).

35. G. Romano, Phonon Transport in Patterned Two-Dimensional Materials from First Principles. arXiv arXiv:2002.08940 [Preprint] (2020). https://doi.org/10.48550/arXiv.2002.08940.

36. V. Chiloyan, S. Huberman, A. A. Maznev, K. A. Nelson, G. Chen, Thermal transport exceeding bulk heat conduction due to nonthermal micro/nanoscale phonon populations. *Applied Physics Letters* **116**, 163102 (2020).

37. K. Dudde, M. Elhajhasan, G. Würsch, J. Themann, J. Lierath, D. Paul, N. H. Protik, G. Romano, G. Callsen, Phonon Mean Free Path Spectroscopy By Raman Thermometry. arXiv arXiv:2505.14506 [Preprint] (2025). https://doi.org/10.48550/arXiv.2505.14506.

36. E. Pop, S. Sinha, K. E. Goodson, "Monte Carlo Modeling of Heat Generation in Electronic Nanostructures" in *Heat Transfer, Volume 7* (ASMEDC, New Orleans, Louisiana, USA, 2002), pp. 85–90.

39. H. Teisseyre, P. Perlin, T. Suski, I. Grzegory, S. Porowski, J. Jun, A. Pietraszko, T. D. Moustakas, Temperature dependence of the energy gap in GaN bulk single crystals and epitaxial layer. *Journal of Applied Physics* **76**, 2429–2434 (1994).

40. W. Shan, A. J. Fischer, S. J. Hwang, B. D. Little, R. J. Hauenstein, X. C. Xie, J. J. Song, D. S. Kim, B. Goldenberg, R. Horning, S. Krishnankutty, W. G. Perry, M. D. Bremser, R. F. Davis, Intrinsic exciton transitions in GaN. *Journal of Applied Physics* **83**, 455–461 (1998).

41. D. Y. Song, S. A. Nikishin, M. Holtz, V. Soukhoveev, A. Usikov, V. Dmitriev, Decay of zone-center phonons in GaN with A1, E1, and E2 symmetries. *Journal of Applied Physics* **101**, 053535 (2007).

42. B. Wei, Y. Li, W. Li, K. Wang, Q. Sun, X. Yang, D. L. Abernathy, Q. Gao, C. Li, J. Hong, Y.-H. Lin, High-order phonon anharmonicity and thermal conductivity in GaN. *Phys. Rev. B* **109**, 155204 (2024).

43. D. Alvarez, JUWELS Cluster and Booster: Exascale Pathfinder with Modular Supercomputing Architecture at Juelich Supercomputing Centre. *JLSRF* **7**, A183 (2021).

44. N. H. Protik, B. Kozinsky, Electron-phonon drag enhancement of transport properties from a fully coupled *ab initio* Boltzmann formalism. *Phys. Rev. B* **102**, 245202 (2020).

**Acknowledgments:**

The authors acknowledge the Gauss Centre for Supercomputing e.V. (www.gauss-centre.eu) for funding this project by providing computing time on the GCS Supercomputer JUWELS (*43*) at the Jülich Supercomputing Centre (JSC). Furthermore, the authors acknowledge J.-F. Carlin for the epitaxy of all III-nitride material used in this study and J. Ordonez-Miranda for valuable discussions.

**Funding:**

N.H.P. and E.T. acknowledge funding from the "Deutsche Forschungsgemeinschaft" (DFG, German Research Foundation) for an "Emmy Noether" research grant (Grant No. 534386252).

G.R. acknowledges funding from the MIT-IBM Watson AI Laboratory (Challenge No. 2415).

G.R. and G.C. acknowledge funding from the MIT Global Seed Funds.

M.E. and G.C. acknowledge funding from the Central Research Development Fund (CRDF) of the University of Bremen for the project "Joint optical and thermal designs for next generation nanophotonics".

The research of M.E., K.D., G.W., J.L. and G.C. was funded by the major research instrumentation program of the DFG (Grant No. 511416444).

The research of I.R. was funded by the Swiss National Science Foundation through Grant No. 200020_162657.

G.C. also acknowledges the MAPEX-CF Grant for Correlated Workflows (Grant No. 40401080) and funding from the MAPEX "Minor Instrumentation Grant" associated to the APF program "Materials on Demand" (MI06/25 and MI07/25).

# Supplementary material to the paper: When heat goes astray - non-local heating in a semiconductor

**Authors:** M. Elhajhasan[1]*, E. Trukhan[3], K. Dudde[1], G. Würsch[1], J. Lierath[1], I. Rousseau[2], R. Butté[2], N. Grandjean[2], N. H. Protik[3], G. Romano[4], G. Callsen[1]*

**Affiliations:**

[1]Institut für Festkörperphysik, Universität Bremen; Bremen, 28359, Germany.

[2]Institute of Physics, École Polytechnique Fédérale de Lausanne (EPFL); Lausanne, 1015, Switzerland.

[3]Institut für Physik and Center for the Science of Materials Berlin (CSMB), Humboldt-Universität zu Berlin; Berlin, 12489, Germany.

[4]MIT-IBM Watson AI Lab, Massachusetts Institute of Technology (MIT); Cambridge, 02141, USA.

*Corresponding author. Email: elhajmah@uni-bremen.de (M.E.); Email: gcallsen@uni-bremen.de (G.C.)

**Abstract:** This document contains the supplementary material (SM) to the paper "When heat goes astray - non-local heating in a semiconductor".

**Supplementary Materials**

Materials and Methods

Supplementary Text

Figs. S1 to S10

Tables S1 to S2

References (*1–18*)

**Materials and methods:**

Setup:

The setup is the fully customized two-laser Raman thermometry (2LRT) + photoluminescence (PL) apparatus that is used in (*1*), which has been further optimized for a high position stability of the sample and the heat and probe laser with respect to each other. Details can be found in (*2*).

Figure S1A shows a simplified sketch of the apparatus. A heat ($\lambda_{heat}$ = 266 nm) and a probe ($\lambda_{probe}$ = 488 nm) laser beam are focused using a single 80x UV microscope objective (numerical aperture = 0.55). The back-scattered signal induced by the probe laser is collected from the sample by the same objective and is guided towards the detection system (monochromator and charged couple device - CCD). To position the heat laser focus spot on a desired position, the objective is moved using an x/y/z piezo actuator. The movement of the probe laser spot around the heat laser spot is enabled by tilting the probe beam at the entrance of the objective. The tilt is achieved by the conjunction of an angular piezo scanner and the two lenses L1 and L2 (4f-system) seen in fig. S1A (not drawn to scale). Additional details can be found in (*1*).

Sample:

The samples used in this study all consist of a stack of III-nitride epilayers (GaN, AlN, $In_{0.15}Ga_{0.85}N$) grown on an n-type (111) silicon substrate. After growth, these epilayers have been etched to create the various geometries presented in this study. Finally, an under-etching was performed to release parts of the epilayer from the Si substrate.

Figure S1B shows the epilayer stack forming the membrane. On the n-type (111) Si substrate, the nitride layers have been grown by metal-organic vapor phase epitaxy (MOVPE). First, 50 nm of AlN were deposited on the substrate to avoid melt-back etching (*3*). Subsequently, 110 nm of GaN, 3 nm of $In_xGa_{1-x}N$ ($x$ = 0.15), and a final layer 90 nm of GaN were deposited. Sandwiched between the two GaN layers, the $In_xGa_{1-x}N$ layer constitutes a quantum well (QW). It has been deliberately placed 10 nm above the geometric center of the neighboring GaN to account for the lower refractive index of GaN. This strongly improves the coupling of the fundamental electromagnetic mode to the QW (*4*), which makes this structure ideal for lasing applications (*1*, *2*, *4*). This membrane has been chosen because it is a frequently studied and fully characterized structure (*1*).

To obtain the structures studied in this paper (see fig. 1C - G), the epilayer needs to undergo electron-beam lithography, dry etching, wet etching, and chemically selective vapor phase etching. A negative tone resist, hydrogen silsesquioxane, is applied on the sample surface. After writing the pattern on the resist and the dry etching process to remove the unexposed areas, a $XeF_2$ chemically selective vapor phase etching is performed to release part of the III-nitrides epilayers from the Si substrate. Additional information and details regarding the growth and the processing of these structure can be found in (*1*) and its supplementary material as well as (*4*).

Figure S1C shows the processed III-nitride pad based on a scanning electron microscope (SEM) image (the inlet shows a light microscope image, visible spots are caused by indium concentration fluctuations in the QW). This geometry consists of a supported membrane, where the edges have an overhang of 8 to 10 µm. Two regions are marked by the roman numerals "I" and "II" (red). These are the regions where the edge ($n$ = 1) and the corner ($n$ = 2) configuration have been measured. Fig. 1B and C shows the measurements performed on the edge ($n$ = 1), while Fig. 2A and B and fig. S5 shows the results from the corner geometry ($n$ = 2). Fig. S2A - C presents a 3D rendering of the pad structure that formed the basis for our COMSOL® simulations.

Fig. S1D-G shows the processed suspended hexagons structures based on SEM images (fig. S1D,F,G) and light microscope image (fig. S1E). These hexagon structures consist of a small 4 µm large hexagon that is suspended by six nanobeams (one on each corner of the hexagon) that are connected to a larger supported hexagon. Arms with lengths *l* = 10, 15, 20, 25, 30, and 35 µm have been fabricated. The larger the length of the arms, the more thermally insulated the inner hexagon becomes. In this study, we present results for *l* = 10, 15, and 30 µm. Fig. S2D presents a 3D rendering of the processed structure showing all relevant geometrical parameters in use for our COMSOL® simulations.

## Numerical methods:

In this section, we will describe the various numerical models we employed in this study. We will start by describing two different trials that we used to reproduce the temperature distributions we observe in our experiments, then we will move on to the description of the calculations of the phonon density of states and the energetically resolved mean free path (MFP) distributions. See (*1*) for details.

To reproduce the experimental temperature distributions, we used both the finite element method (FEM) software COMSOL® and the open-source OpenBTE python package (*5*, *6*). The FEM software was used to solve the heat equation on numerical replicas of the structures we measured, while the OpenBTE package solves the Boltzman transport equation (BTE) on 2D models of the structures we probed.

Lastly, the purpose of the density of states and energetically resolved MFPs is to shed light on the mechanisms that lead to the appearance of the edge heating phenomenon. In the main text, we explain its appearance by the presence of phonons with large MFP values and the optical heating of the sample. These calculations support our interpretation of the edge heating by showing that such phonons exist and that an optical heating can generate them at elevated temperatures.

### *General simulation parameters:*

When creating the various numerical models to reproduce the experimental results, several parameters have to be determined during the experiments or have to be taken from literature. These are spot sizes of the focused heat and probe laser determined by the knife edge method (*1*), defined as the full width half maximum (FWHM) of their intensity profile, the heat laser power absorbed by the sample, and the penetration depth of the heat laser in the sample.

The spot sizes of the heat and probe laser are equal to $FWHM_{heat}$ = 500 nm and $FWHM_{probe}$ = 1200 nm, respectively. The penetration depth of the heat laser into GaN is equal to $p_{heat}$ = 45 nm. The power absorbed by the sample is measured for each experiment and the corresponding value are given in the inset of each map $T_{ph}(x,y)$ on each temperature mapscan.

### *COMSOL® software model:*

#### *Supported GaN-based square membrane*

Fig. 2A and B shows the numerical structure corresponding to fig. S1C. The yellow block corresponds to the Si support, while the blue thin top layer corresponds to the III-nitride membrane. The III-nitride epilayer has been modelled by a single effective 250-nm-thick GaN layer.

To solve the stationary heat equation, the only material-dependent variable needed is the thermal conductivity $\kappa$. The equation to solve reads as:

$$-\nabla \cdot (\kappa \nabla \mathrm{T}) = Q(x, y, z)$$

with $Q(x, y, z)$ being the function modeling the heat source. For Si, $\kappa$ is set to a constant value of 130 $Wm^{-1}K^{-1}$. For GaN $\kappa$ can either be a constant value $\kappa_0$ = 95 $Wm^{-1}K^{-1}$ that was determined in our previous study on this membrane at room temperature (*1*), or temperature dependent $\kappa$(T) [extracted from (*7*)]. In this case $\kappa$(T) needs to be scaled so that $\kappa(T = 295\ \mathrm{K}) = 95\ Wm^{-1}K^{-1}$ in agreement with our experiments (*1*). Both cases are presented in this study.

The boundary conditions are set to "Thermal insulation" everywhere except at the bottom of the Si support and under the heat laser spot. The bottom of the Si support is set to be equal to the ambient temperature during the measurement. In this study, this value is set to 295 K for the III-nitride pad.

Since the intensity of the laser focus spot follows a Gaussian function, it radiates its energy over the whole top surface, which is why we mostly pursued this implementation. In some cases, it can be beneficial to not introduce the heat source over the whole surface. In this case, a cylinder centered around the heat laser spot location with a radius of 1.2 µm already comprises more than 99.7% of the total power. In addition to the Gaussian profile of the intensity of the focused heat laser spot in the plane perpendicular to its propagation direction, there is an additional exponential reduction of the intensity along the depth of the membrane, due to the absorption of the laser by the sample (Beer-Lambert absorption). Taking everything into account, the heat source is modeled by the following function:

$$Q(x, y, z) = P_{abs} \cdot \frac{e^{z-z_0/p_{heat}}}{A} \cdot \frac{1}{2\pi \cdot \sigma^2} \cdot e^{-((x-x_0)^2+(y-y_0)^2)/2\cdot\sigma^2} \qquad (1)$$

with $P_{abs}$ the power absorbed by the sample, ($x_0$, $y_0$, $z_0$) the coordinates of the focused heat laser spot in cartesian coordinates, $A = p_{heat} \cdot \left(1 - e^{-250\,[nm]/p_{heat}}\right)$ a normalization factor, and $\sigma = FWHM_{heat}/2\sqrt{2ln(2)}$. The integral of this function over the whole membrane is equal to $P_{abs}$.

The mallow circle in fig. S2C shows the location of the heat spot in the edge geometry, which is the experimental situation indicated by the red symbol "I" in fig. S1C.

*Transition area*

To release part of the epilayer from the Si substrate, the sample underwent a vapor phase etching process using $XeF_2$.

The III-nitride membrane partially protects the Si from being etched, therefore, at the base of the support and at the junction of the III-nitride membrane with the Si support, a fillet appears. This has an impact on the final simulated temperature distribution in the membrane, thus, the dimensions of this fillet have to be properly estimated.

We call the area above this fillet the transition area and it is depicted in fig. S5B. The extent of this area has been determined through the analysis of optical microscopy images and SEM images, together with the analysis of unheated Raman and PL mapscans recorded at the edge ("I" in fig. S1C) and the corner ("II" in fig. S1C) of the sample.

The optical images showed a sharp contrast between the freestanding and the supported membrane at a distance of roughly 8 µm from the edge. A similar contrast can be seen at this distance, when analyzing unheated Raman mapscans of the edge or the corner of the membrane [see, e.g. Fig 3a in (*1*)]. Electron microscopy images [e.g. Fig. 1c in (*1*)] show a fillet at the corner at the bottom of the Si substrate, while PL mapscans [e.g. Fig. 2b in (*1*)] show a region of decreased intensity about 8 to 10 µm away from the edges.

The depth of the under-etch was set to 10 µm, therefore its extent below the membrane will also measure 10 µm. Taking this into account, we conclude that at the connection between the GaN-based membrane and the Si substrate, a Si fillet with a curvature radius of 2 µm formed. Therefore, we add it to the numerical models (fig. S2B, transition area labeled with "etching front").

*The suspended hexagon*

Figure S2D shows a schematic of a suspended hexagon. In this geometry, the transition area did not change the result of the simulation in the area of interest, therefore we did not need to model it.

All geometrical parameters needed to reproduce the hexagon structure can be found in fig. S2B and D. Similar to the GaN-based pad, the blue part designates the GaN membrane while the yellow part is the Si substrate. The two black dashed hexagons in the middle of the outer hexagon indicate the location of the Si below the membrane.

*Convolution with the probe beam laser spot*

To compare the simulations with the experiments, it is necessary to convolute the result of the simulations with the modeled probe laser intensity. The simulated probed temperature is given by [SSec. IV D in (*1*)]:

$$T_{probe}(\vec{r}_0) = \frac{1}{p_{probe} \cdot w_e^2 \cdot \pi} \int_V dr\, d\theta\, dz \cdot r \cdot T(r,\theta,z) \cdot e^{-\frac{(\vec{r}-\vec{r}_0)^2}{w_e^2}} \cdot e^{-\frac{2z}{p_{probe}}} \tag{2}$$

where $p_{probe}$ is the probe beam penetration depth and the factor 2 in the exponent of the exponential function accounts for the probe beam going in and out of the material, $w_e^2$ is the laser beam waist, related to the beam FWHM through $w_e = \frac{FWHM}{2\sqrt{ln(2)}} \cong \frac{FWHM}{1.66}$. The parameters $r, \theta$ and $z$ are the cylindrical coordinates centered at the top surface of the membrane in the middle of the heat spot, and $\vec{r}_0$ is the center of the probe beam in cylindrical coordinates.

*OpenBTE numerical model:*

We enhance the numerical analysis of the experimental results presented in this study by solving the single MFP two-dimensional phonon BTE. The simulation, being two dimensional, allowed us to simplify the geometry which can be stopped at the location of the Si substrate. Only one material is considered, GaN, and the Si substrate is modelled by thermostating the outer edge of the geometry to the ambient temperature in the cryostat. For the results presented in Fig. 3A and B, this value was set to 12 K.

*Boltzmann Transport Equation*

The following formalism describes the modelling behind Fig. 3C and D in the main text. The bulk material is described by a single MFP Λ and a bulk thermal conductivity $\kappa$, which we set to 200 $Wm^{-1}K^{-1}$ (more details in the Supplementary text on page 9). Within this simple model, phonon modes are parametrized by the polar angles $\phi_\mu$, where $\mu = [0, ..., N - 1]$, with $N = 72$. The BTE then reads:

$$\beta \Lambda \hat{\mathbf{s}}_\mu \cdot \nabla \mathrm{T}_\mu(\mathbf{r}) + \beta \mathrm{T}_\mu(\mathbf{r}) = \frac{\beta}{N} \sum_{\mu'} \mathrm{T}_{\mu'}(\mathbf{r}) + \frac{1}{N} Q(\mathbf{r}) \tag{3}$$

where Λ [m] is the MFP, $\hat{\boldsymbol{s}}_\mu = \sin(\phi_\mu)\cdot\hat{\boldsymbol{x}} + \cos(\phi_\mu)\cdot\hat{\boldsymbol{y}}$ is the phonon direction, $\beta = 2\kappa/\Lambda^2$, and Q(r) is the space-dependent heat source [$Wm^{-2}$], which has the Gaussian profile:

$$Q(\mathbf{r}) = \frac{Q_{tot}}{2\pi\sigma^2} \exp\left(-\frac{|\boldsymbol{r}|^2}{\mathbf{2}\sigma^2}\right) \tag{4}$$

with $Q_{tot} = 0.86$ mW, and $\sigma = FWHM/2\sqrt{2ln(2)}$, with FWHM = 500 nm. These values match the experimental parameters that were used for the acquisition of Fig. 3A and B. The unknown $\mathrm{T}_\mu(\mathbf{r})$ are the deviation of the phonon effective temperature from a reference temperature $T_0 = 0$ K. Within this formalism, the effective temperature and heat flux are given by:

$$\mathrm{T}(\mathbf{r}) = \frac{1}{N} \sum_{\boldsymbol{\mu}} \mathrm{T}_{\boldsymbol{\mu}}(\mathbf{r}) \tag{5}$$

$$\mathbf{J}(\mathbf{r}) = \frac{2\kappa}{\Lambda N} \sum_{\boldsymbol{\mu}} \hat{\mathbf{s}}_\mu \mathrm{T}_{\boldsymbol{\mu}}(\mathbf{r})\,. \tag{6}$$

*The boundary conditions*

- Thermostated boundary conditions are enforced by setting $T_\mu = T_b$ for incoming phonons.
- Adiabatic boundaries are applied via reflection coefficients acting on the incoming temperature distributions:

$$\mathrm{T}_\mu = \sum_{\boldsymbol{\mu}'} \Theta_\mu^- R_{\mu\mu'} \Theta_{\mu'}^+ \mathrm{T}_{\boldsymbol{\mu}'} \tag{7}$$

  where $\Theta_\mu^\pm = H(\pm\hat{\mathbf{s}}_\mu \cdot \hat{\mathbf{n}})$, $\hat{\mathbf{n}}$ is the normal to the boundary, and $H(\ldots)$ is the Heaviside function.

  - For diffuse scattering we have

$$\mathrm{R}_{\mu\mu'}^D = \frac{\hat{\mathbf{s}}_{\mu'} \cdot \hat{\mathbf{n}}}{\sum_{\mu''} \hat{\mathbf{s}}_{\mu''} \cdot \hat{\mathbf{n}}\, \Theta_{\mu''}^+} \tag{8}$$

  - For specular reflection we have:

$$\mathrm{R}_{\mu\mu'}^S = \delta\left(\hat{\mathbf{s}}_\mu - \hat{\mathbf{s}}_{\mu'} + 2(\hat{\mathbf{s}}_{\mu'} \cdot \hat{\mathbf{n}})\hat{\mathbf{n}}\right). \tag{9}$$

These boundary conditions all lead to $\mathbf{J} \cdot \hat{\mathbf{n}} = 0$. Additionally, the diffuse boundary conditions BC leads to zero tangential flux of the phonons leaving the surface:

$$\sum_{\boldsymbol{\mu}} \Theta_\mu^- \hat{\mathbf{s}}_\mu \mathrm{T}_{\boldsymbol{\mu}} = \sum_{\boldsymbol{\mu}} \hat{\mathbf{s}}_{\mu,t} \mathrm{T}_{\boldsymbol{\mu}} = \mathbf{0} \tag{10}$$

where $\hat{\mathbf{s}}_{\mu,t}$ is the tangential component of the flux in the $\mu$ direction. Consequently, the overall flux close to the edges is expected to be smaller in the diffusive case (relative to the overall flux) than in the specular case. Equation 3 is discretized using the finite-volume upwind scheme implemented in OpenBTE (*5*, *6*).

*The convolution with the probe beam laser spot*

In a two-dimensional simulation, Eq. 2 reduces to:

$$T_{probe} = \frac{1}{w_e^2 \cdot \pi} \int_S dr\, d\theta \cdot r \cdot T(r,\theta) \cdot e^{-\frac{\left(\vec{r}-\vec{r}_0\right)^2}{w_e^2}}. \tag{11}$$

Because the 2D simulations were performed using OpenBTE, no built-in function can evaluate integrals on a given domain. Therefore, we took the result of the simulation, which are the set of discrete temperature values located at the nodes of the mesh, and convoluted them with the 2D probe laser intensity function:

$$T_{probe}(\vec{r_0}) = \frac{1}{F_{norm}} \sum_{i,j} T_{i,j} e^{-\frac{\left(\vec{r}_{i,j}-\vec{r}_0\right)^2}{w_e^2}}. \tag{12}$$

$$F_{norm} = \sum_{i,j} e^{-\frac{\left(\vec{r}_{i,j}-\vec{r}_0\right)^2}{w_e^2}} \tag{13}$$

with $\vec{r}_0$ being the position of a node in the mesh. The operation is repeated for every point $\vec{r}_0$ belonging to the mesh. $\vec{r}_{i,j}$ and $T_{i,j}$ are the position and the temperature of the mesh node with index (*i,j*), respectively, and the sum is made over every point in the mesh.

The resulting $T_{probe}(\vec{r}_0)$ values are then imported into a graphing software to plot the simulated distributions and extract cutlines as shown in Figs. 3C, D and S9A-K.

<u>Density of states and energetically resolved MFP distributions</u>

*The density of states*

The phonon density of states is defined as:

$$g(\omega) = \frac{1}{N} \sum_{s\mathbf{q}} \delta\left(\omega - \omega_{s\mathbf{q}}\right) \tag{14}$$

where $\omega_{s\mathbf{q}}$ is the angular frequency of the phonon mode characterized by wave vector $\mathbf{q}$ and branch s and N is the number of primitive unit cells in the crystal. Note that this is normalized such that $\int d\omega\, g(\omega) = N_s$, where $N_s$ is the number of phonon branches (thrice the number of atoms in the primitive unit cell). For the data in Fig. 3E of the main text, we evaluate the above equation on-shell, i.e., for $\omega = \omega_{s\mathbf{q}}$. We used the analytical tetrahedron method of Lambin and Vigneron (*8*) to evaluate the energy conserving δ-function. In Fig. 3E, we plotted the density of states scaled by the equilibrium phonon distribution, i.e., the Bose-Einstein function:

$$n^0\left(\omega_{s\mathbf{q}}\right) = \left[\exp\left(\hbar\beta\omega_{s\mathbf{q}}\right) - 1\right]^{-1} \tag{15}$$

where $\hbar$ is the reduced Planck's constant and β is the inverse temperature energy. The plotted quantity $g(\omega_{s\mathbf{q}})n^0(\omega_{s\mathbf{q}})$ is the spectrum of the thermal phonon occupancy per unit cell volume.

*Phonon mean free paths at 300 K*

The mode-resolved phonon mean free paths (MFP) vs. mode energies (shown in fig. S10) was calculated using **elphbolt** (*9*). The MFP was defined using the following formula:

$$MFP_{s\mathbf{q}} \equiv \frac{\mathbf{F}_{s\mathbf{q}} \cdot \hat{\mathbf{v}}_{s\mathbf{q}}}{k_B} \quad (16)$$

where $\hat{\mathbf{v}}_{s\mathbf{q}}$ is the unit vector of the phonon group velocity, $k_B$ is the Boltzmann constant, and $\mathbf{F}_{s\mathbf{q}}$ is the phonon response function to the applied field $\nabla_{\mathbf{r}}T$, obtained by iteratively solving the linearized phonon BTE (*9*) beyond the relaxation time approximation. Note that $\mathbf{F}_{s\mathbf{q}}/k_B$ as defined in the **elphbolt** theory, has units of length and can be interpreted as a mean-free-displacement, i.e., a vectorial generalization of the MFP (*10*). In this work, we include the following interactions affecting the transport properties of the phonon system: three- and four-phonon interactions, phonon-thin-film scattering, and phonon-isotope scattering.

The phonon BTE solver requires input data generated from density functional and density functional perturbation theories. These include the second-, third-, and fourth-order interatomic force constants. Furthermore, for the inclusion of 4-phonon scattering in **elphbolt**, we calculated the corresponding scattering rates using the **FourPhonon** code (*11*) on a 12 × 12 × 12 wave vector mesh and then trilinearly interpolated to the finer converged mesh . Details of the computational workflow of these calculations have previously been given in (*1*).

For the phonon-thin-film scattering rates, we set the film width $w$ to 250 nm and used the phenomenological expression $2\mathbf{v}_{s\mathbf{q}}^{\perp}/w$, where $\mathbf{v}_{s\mathbf{q}}^{\perp}$ is the group velocity component normal to the thin-film. The phonon-isotope scattering rates were computed using two approaches: the Tamura model (first Born approximation on top of the virtual crystal approximation) (*12*) and the DIB-1B method (first Born approximation atop the dominant isotope background method) (*13*). The dependence of MFPs on phonon energies and the resulting thermal conductivities for these two approaches are presented in fig. S10 and table S2, respectively. One can see that both approaches yield similar results. In this work, we employed the more recently proposed DIB-1B theory.

<u>Temperature calibration:</u>

To deduce temperatures from the $E_2^{high}$ Raman mode shift $\omega$ and broadening $\Delta\omega$, a temperature calibration of both is necessary. Such a calibration is displayed in fig. S3. The complete procedure detailing the extraction of the temperature from $\omega$ and $\Delta\omega$ is explained in (*1*).

We found that the Stokes/anti-Stokes intensity ratio of Raman modes, which can also be used to derive temperatures without the need for any temperature calibration, was not suitable for any mapping of $T_{ph}(x,y)$ in our membrane structures. In addition to the long integration times and large uncertainties (*14*) of this method, our III-nitride epilayer is a photonic membrane which is meant to emit light. For suitable probe wavelength (here 488 nm), any anti-Stokes signal is buried in PL signal coming from the QW.

**Supplementary text:**

Comparison of the experimental data with simulations – cutlines parallel & perpendicular to the edge ($n$ = 1):

In addition to the cutlines perpendicular to the edge ($n$ = 1) of the membrane (Fig. 1E and F), one can also analyze the cutlines parallel to the edge. Figure S4 displays such cuts. In fig. S4B, where $d$ = 4.6 μm, both experimental (red-connected squares) and convoluted simulated (black connected dots) curves agree with each other. In fig. S4C, where $d$ = 1.9 μm, the anomalous heating at the edge reappears and the experimental temperature exceeds its simulated counterpart by up to 50 K. Fig. S4B and C additionally display the non-convoluted simulated curve (black lines) to show the impact of the convolution on the temperature distribution.

When comparing Fig. 1E and F one notices a non-zero $T$-slope at the edge of the membrane for $d$ = 1.9 μm, despite the surface at the edge being adiabatic. One could interpret this behavior as being in stark contrast with Fourier's law. Romano *et al.* show (*15*) that this effect can be approximated by an effective Robin-type boundary condition derived from a coarse-graining of the BTE, where the boundary conductance is computed from the MFP distribution. In the presence of strong $T$-gradients, a Fourier simulation (convoluted with the probe beam intensity profile) can also show a non-zero $T$-slope at the edge of the membrane. Therefore, using such a $T$-slope criterion as a signature for non-Fourier phonon transport needs a detailed comparison of (non-linear) Fourier simulations to the experimental data.

Confirmation of the anomalous heating at the edge by another temperature sensor:

As the relative energetic position $\omega$ of the $E_2^{high}$ Raman mode of GaN is sensitive to strain, one could argue that the anomalous additional heating at the edge is due to a misinterpretation of the mode shift (*1*). This explanation does not hold under the scrutiny of the shift of the $E_2^{high}$ mode under thermally induced strain (*14*), nor explains the presence of the anomalous heating when looking at temperature maps based on the broadening of the $E_2^{high}$ mode $\Delta\omega$. Figure S5A shows such a temperature map where edge heating is visible in the corner geometry ($n$ = 2). Here the noise in the $T(x,y)$ map based on $\Delta\omega$ originates from the limited signal to noise ratio in the associated Raman spectra. Figure S6A shows a temperature map based on the broadening $\Delta\omega$ of the $E_2^{high}$ Raman mode for the hexagon geometry ($n$ = 6) where edge heating is unequivocally visible. Figures S5A and 2B originate from the same measurement, and so do Figs. S6A and 3A.

Figure S6B shows spectra from the position marked by "i, ii and iii" in Fig. 2F in the main text. For each location two spectra from adjacent points are shown with their respective fits. We employ a Voigt function to fit our spectra. The gaussian part of the Voigt function is determined by fitting the width of a mercury spectral lamp with a gaussian function and subsequently used as a fixed parameter in the Voigt fit. Only the Lorentzian part of broadening becomes a fitting parameter. See (*1*) for more details on the fit function.

One clearly sees in fig. S6B that the Raman mode both broadens and red-shifts at the edge, indicating an increase in temperature.

Onset of the edge heating:

Figure S5B shows the temperature distribution for $n$ = 2 and $P_{abs}$ = 4.03 mW. The onset of edge heating can be seen at both edges of the membrane. This recording adds to the series of the two $T(x,y)$ maps presented in Fig. 2A and B.

Comparing this series ($n$ = 2) with the measurements taken at the edge ($n$ = 1) and at the hexagons ($n$ = 6) hints toward a phenomenon appearing at elevated temperature, rather than a phenomenon triggered by higher laser powers. In the case of the edge, even a power absorbed as high as 8.07 mW (Fig. 1B) is not enough to clearly show the edge heating phenomenon, while for the hexagons a power absorbed as low as 0.86 mW is already enough to exhibit the edge heating phenomenon, as long as the temperature rise is above a certain temperature (approx. 500 K). Fig. S7 shows the result of a 2LRT measurement with an absorbed power of 1.2 mW on a suspended hexagon with $l$ = 10 μm (Figs. 2E and S7 form a set). The ambient temperature was set to 295 K and the hexagon ($n$ = 6) reached a maximum temperature of 570 K, similar to Fig. 2B ($n$ = 2). Temperature-wise both measurements depicted in Figs. 2B and S7 started at the same base temperature and ended at a comparable one, yet only the measurement in Fig. 2B manifested an edge heating phenomenon. The interplay of the absolute temperature, the temperature gradient, the proximity of the boundaries and $P_{abs}$ can all influence the presence of the edge heating phenomenon. The absolute temperature rise and the optical injection of heat seem to be the predominant elements for the appearance of the edge heating.

Comparison of the experimental data with simulations – hexagon geometry ($n$ = 6):

We present the numerical analysis of the hexagon geometry with $l$ = 30 μm, $P_{abs}$ = 0.86 mW, and an ambient temperature of 12 K. These are the parameters of the 2LRT measurement shown in Fig. 3A and B. We simulated three types of models: the first is a heat-equation-based model with a constant thermal conductivity $\kappa_0$ of 95 $Wm^{-1}K^{-1}$ (*1*), the second considers a temperature-dependent thermal conductivity coefficient $\kappa$(T), and the last is a 2-dimensional simulation of the Boltzmann transport equation (BTE), with a single MFP. The simulation with $\kappa$(T) yielded temperature rise values close to the measured ones (fig. S8C) and the BTE simulations were able to reproduce the edge heating effect in the hexagon structure (Fig. 3C and fig. S9) based on the three criterion we exposed in the main text F1-F3.

*Simulation with a constant thermal conductivity $\kappa_0$:*

This simulation was performed using the FEM software COMSOL®. To match the measurement shown in Fig. 3A and B, the power absorbed by the sample is set to 0.86 mW and is introduced as a volumetric heat source in the center of the hexagon, modelled by Eq. (1). The ambient temperature was set to $T_{amb}$ = 12 K.

One notices that the central hexagon is supported by six nanobeams, measuring 0.5 μm in width, 0.25 μm in height, and 30 μm in length. Similar to the reduction of the thermal conductivity that occurs when one compares a bulk sample with a thin membrane (*3D → 2D* transition), another reduction happens when one carves beams from a membrane (*2D → 1D* transition). For a first rough qualitative agreement between measurements and simulations, we applied the same reduction that we observed for the bulk to membrane transition, to the beam situation.

Mathematically: $\kappa_{membrane} = \alpha \cdot \kappa_{bulk} \rightarrow \kappa_{beam} = \alpha \cdot \kappa_{membrane} = \alpha^2 \cdot \kappa_{bulk}$ *(3D→2D→1D)*. In our case, we took a value of 200 $Wm^{-1}K^{-1}$ (*7*, *16*) for $\kappa_{bulk}$ of GaN. With $\kappa_{membrane} = 95$ $Wm^{-1}K^{-1}$ measured in (*1*), we deduce $\kappa_{beam} = 45.125$ $Wm^{-1}K^{-1}$, and α = 0.475. In literature, the value of $\kappa_{bulk}$ varies from 150 $Wm^{-1}K^{-1}$ to more than 250 $Wm^{-1}K^{-1}$ (*7*, *16*), depending on the sample quality. Thus, we took the average value of 200 $Wm^{-1}K^{-1}$.

This simulation failed both qualitatively and quantitatively (fig. S8C, grey dotted line). Qualitatively it failed due to the Fourier equation itself. In this picture, the flow of heat is linked to the gradient of temperature. As such, the heat source will always be the hottest spot and no edge heating can be observed in this configuration. The quantitative failure can be attributed to the extreme temperature gradients (870 K to 12 K) present in the sample when considering the full path from the heat source to the heat sink. Between 12 K and 870 K, the thermal conductivity value spans from ≈ 50 $Wm^{-1}K^{-1}$ at high temperatures (*7*) to more than 1000 $Wm^{-1}K^{-1}$ at low temperatures around 50 K (*16*) for bulk GaN samples.

*Simulation with a temperature-dependent thermal conductivity κ(T)*

The first step to improve the model was to introduce the temperature dependence of the thermal conductivity $\kappa$(T). Its trend has been extracted from (*7*) and was scaled down to correspond to our membrane at $T$ = 295 K: $\kappa(\mathrm{T}) = 0.475 \cdot \left(11.87 + \frac{526.5-11.87}{1+\left(\mathrm{T}/_{216.67}\right)^{1.85}}\right)$. It should be noted that this function works well until a bit below 100 K. At around 100 K, the temperature dependence of the thermal conductivity of wurtzite GaN stops following the $T^{-1}$ power law and has a maximum around 50 K before sharply declining to 0 $Wm^{-1}K^{-1}$ toward 0 K (*16*). Since the whole hexagon and its arms have temperatures well above 100 K, this is not of great concern for our comparison with the experiment.

In fig. S8C, the red dotted line shows the comparison of the experiment with the simulation based on $\kappa(T)$. The shape of the simulated temperature rise does not match its measured counterparts, yet the overall temperatures are in the same range. In the arms, the measured temperature is larger than its simulated counterpart. This is explained by the inadequacy of the thermal conductivity function at low temperatures. Since it overestimates the thermal conductivity of wurtzite GaN at low temperatures, the simulated support structure around the suspended hexagon becomes a better heat dissipator in the simulation than its real counterpart, thus explaining the lower temperatures in the simulated arms.

*2D Boltzmann equation with a single phonon mode*

Fig. S9 shows the result of the 2D BTE simulations. The data are compared through cutlines, equivalent to the ones used in fig. S8 (cutlines along the supporting arms). Fig. S9A-J displays the normalized convolution of the result of the simulations (MFP = 8-80 μm + Fourier) with our probe laser profile. Thus, these temperature profiles consider the resolution limit of the present 2LRT setup.

In these simulations, the edge heating phenomenon occurs at fairly large MFP values (> 32 μm), while experimentally, we expect the effect to start at MFP values around 2 μm. We explain this discrepancy, i. a., by the inability of the current model to account for the full spectrum of MFP values present in our material (see Fig. 3E and F for comparison), as well the linearized version of the BTE that we are employing, which is not well suited to study phenomenon displaying large temperature gradients. The scattering of a long MFP phonon can generate any allowed phonon combinations in the system (details follow), which are likely to have lower MFP values. When a large MFP phonon reaches a boundary and scatters, the resulting lower MFP phonons cannot travel far from the boundary, causing an accumulation of the thermal energy at the boundary. But our theoretical model enforces all phonons to have the same MFP value, failing to properly capture the aforementioned effect. A microscopic description of phonon scattering at a boundary goes beyond the scope of this work. Future theoretical work is needed to improve our present model, which, however, also needs to consider the particular phonons distribution that is evoked by our optical heating, together with non-linear effect occurring when large temperature gradients are present. Such endeavor exceeds the present, mostly experimental study.

Lastly, we justify the 2D nature of the simulation by the extremely flat interfaces of the structure, where the top and bottom interfaces have a root mean square roughness of respectively 1.4 nm and 0.6 nm (*1*). On these interfaces, we assume specular boundary conditions, such that phonons conserve the in-plane component of the momentum upon reflection. In this case, the in-plane heat flux is effectively two-dimensional. On the side interfaces, although the etching process leaves a comparable flat surface behind (*1*), it also introduces defects that act as scattering centers, which justifies the use of diffusive boundary conditions.

## Fig. S1

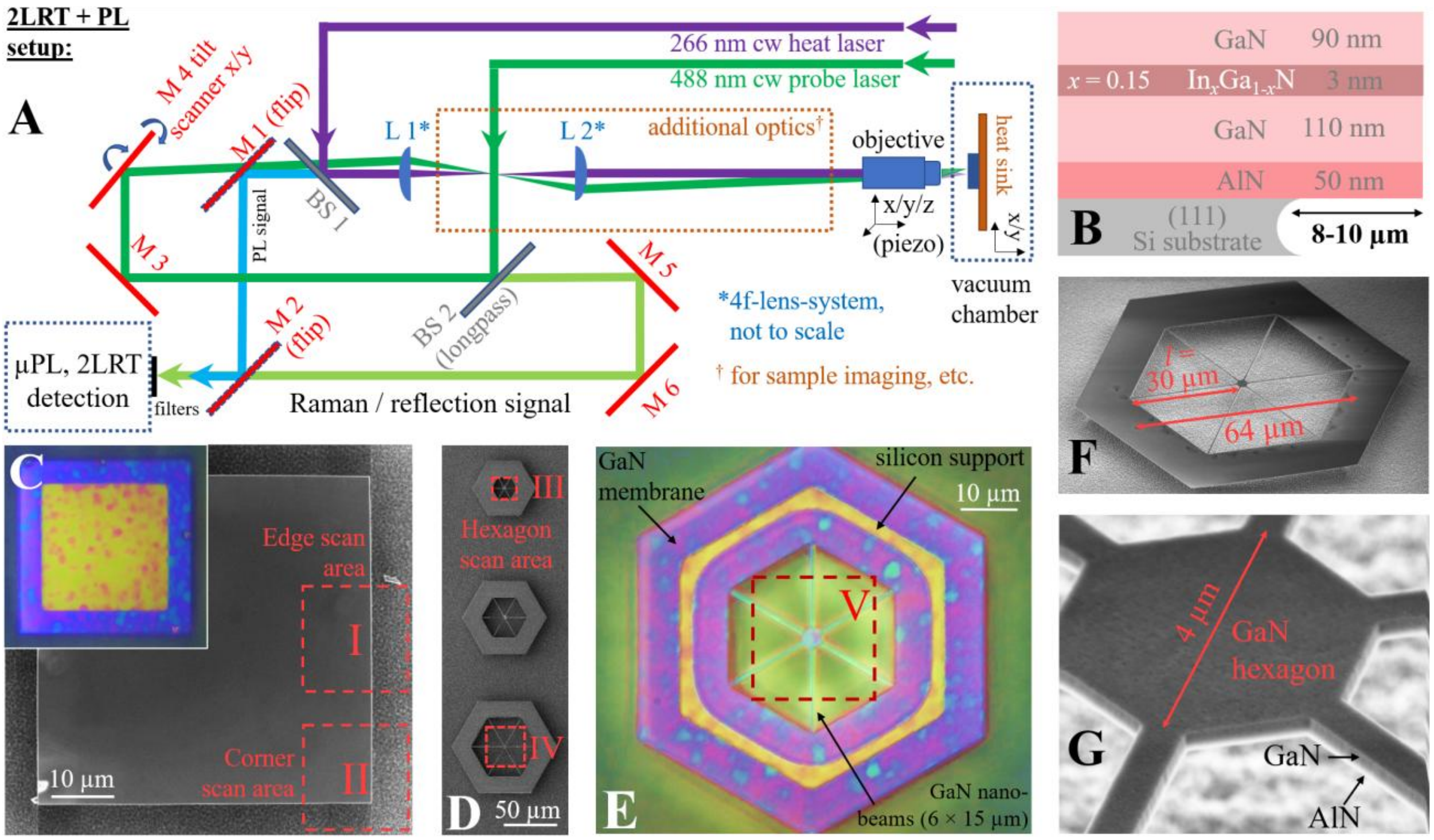

**Fig. S4. 2LRT + PL Setup and sample.** (**A**) Simplified scheme of the optical setup used for the 2LRT + PL experiments. **(B)** Sketch of the layer stack forming the membrane. **(C)** Top-view scanning electron microscopy (SEM) image of the pad where the edge (I) and corner (II) configurations were probed by 2LRT. The inset image is an optical microscope image of a similar pad structure 53.2 x 53.2 μm$^2$ in size. The color difference between the freestanding (blue/purple) and the supported (yellow) parts is created by the different bottom interfaces (membrane/vacuum vs membrane/Si). **(D)** Top view SEM image of three different hexagons suspended by nanobeams with $l$ = 10 μm, 20 μm, and 30 μm. **(E)** Top-view optical image of an $l$ = 15 μm suspended hexagon. The contrast in color due to the different bottom interfaces is again visible. **(F)** Overview SEM image of a suspended hexagon ($l$ = 30 μm). **(G)** Close-up image of **F**. The GaN and AlN layers are visible.

**Fig. S2**

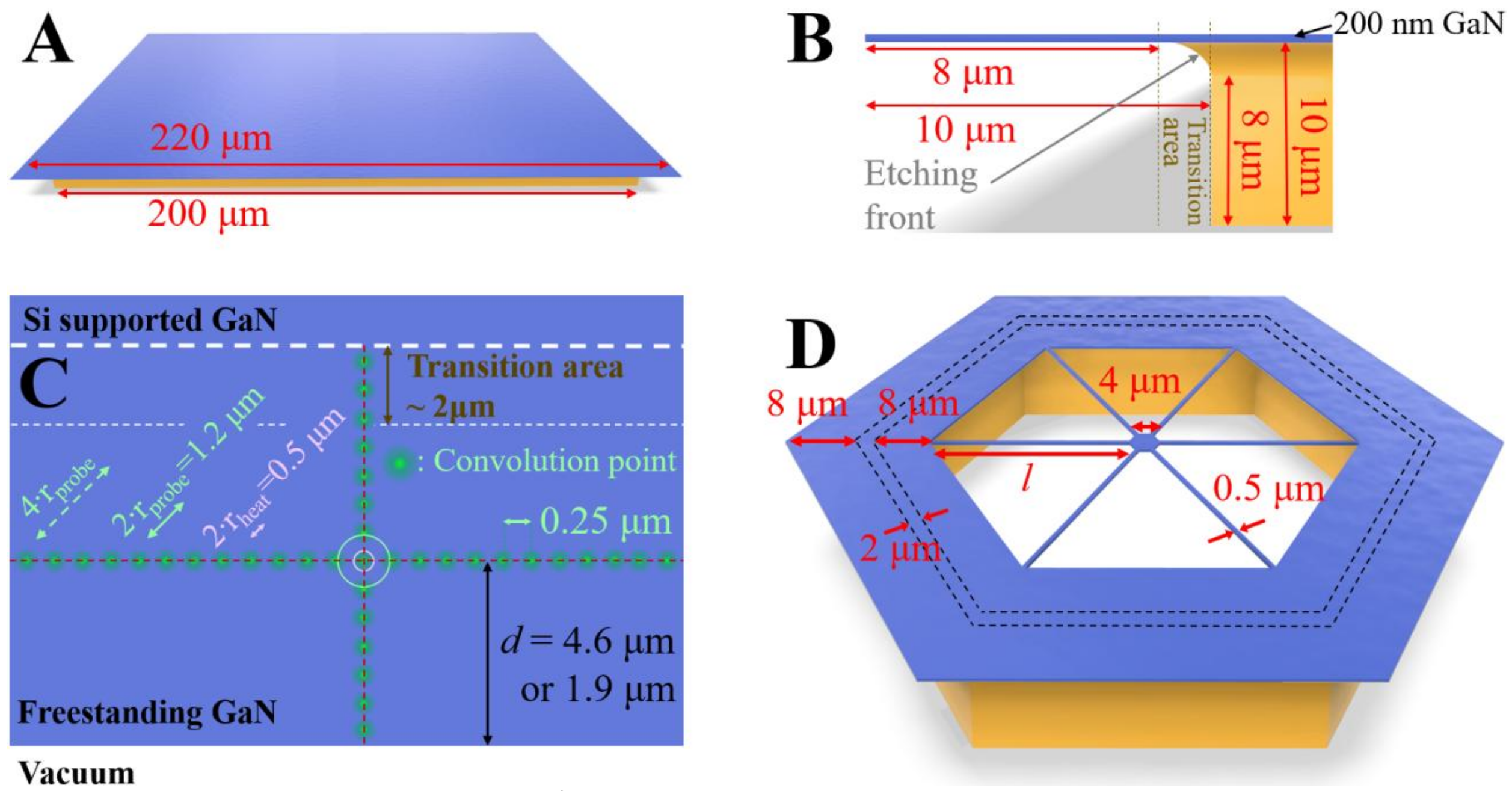

**Fig. S5. Numerical models in COMSOL®.** (**A**) Overview of the pad structure where the temperature distribution at the edge ($n$ = 1) and the corner ($n$ = 2) have been measured. (**B**) Side view of the pad at the edge of the structure, y-axis not to scale. Even if $XeF_2$ etches the Si isotropically in the main crystal directions [sec. II A of (1)], the junction between the Si substrate and the membrane forms a fillet, which is here called the transition area. (**C**) Top view of the membrane for a measurement at its edge. The mauve and green full circles represent the location of the heat and probe laser spots when they overlap, before the probe laser scans the membrane surface. The convolution points are the points where the output of the simulations was convoluted with the probe laser intensity profile to get a value of the temperature comparable to the experiments. The red dotted lines correspond to the cutlines in Fig. 1E and F as well as fig. S4, where the experimental and simulated data are compared. (**D**) Overview of the hexagon structure ($n$ = 6). The distance $l$ is either 10 µm, 15 µm, or 30 µm in this work. The region of the membrane between the two dashed hexagons is supported by Si, while the rest is freestanding. All III-nitride membranes have the same thickness of 250 nm.

**Fig. S3**

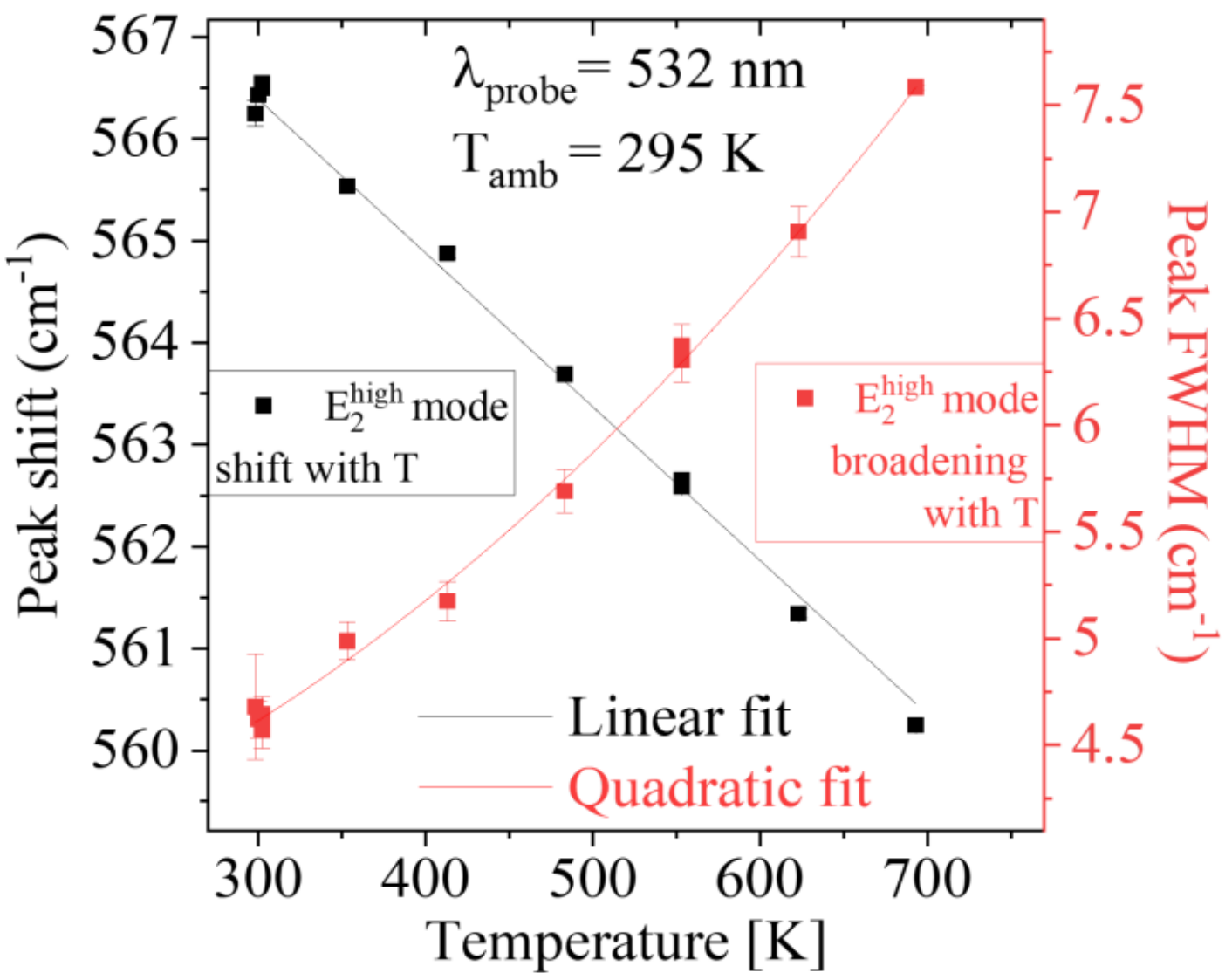

**Fig. S6. Temperature calibration.** (**A**) Temperature calibration of the GaN membrane measured at a corner (probe laser is positioned 4 μm away from each edge) of the pad structure. The $E_2^{high}$ mode position and FWHM have been measured both as the temperature in the heat stage was first increased and subsequently reduced, aiming to detect any potential hysteresis in the system. A similar calibration for bulk GaN with its corresponding spectra can be found in (1). The Raman mode shift $\omega$ and the corresponding FWHM $\Delta\omega$ can be approximated by a linear, respectively quadratic function (solid black and red lines).

## Fig. S4

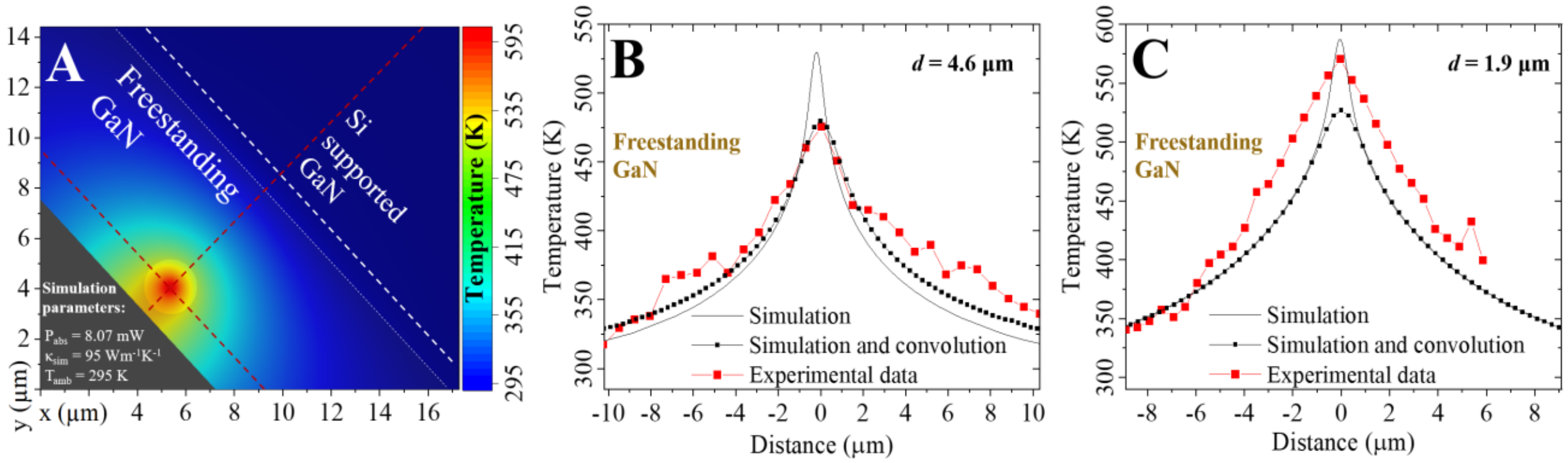

**Fig. S7. Additional temperature profile along the edge (*n* = 1).** (**A**) Simulation of the measurement at an edge, with a distance *d* = 1.9 µm. The red dotted lines correspond to the locations where the experimental and simulated data are compared. The area between the two white dotted lines is the transition area, depicted in fig. S2B. (**B**) Comparison of the experimental and simulated data from a cutline parallel to the edge of the membrane (heat laser spot distance to the edge *d* = 4.6 µm). The continuous black line corresponds to the simulated sample temperature, while the connected black dots represent the simulated sample temperature convoluted with the intensity profile of the probe laser spot. These convoluted temperatures achieve a good agreement with the temperatures that we measure during our 2LRT experiments. The connected red data points correspond to the experimental data. (**C**) Same as **B** but with *d* = 1.9 µm. Similar to Fig. 1F from the main text, one observes a larger discrepancy between the experimental and theoretical data. Error bars have been omitted for clarity (they match the error bars given in Fig. 1E and F).

## Fig. S5

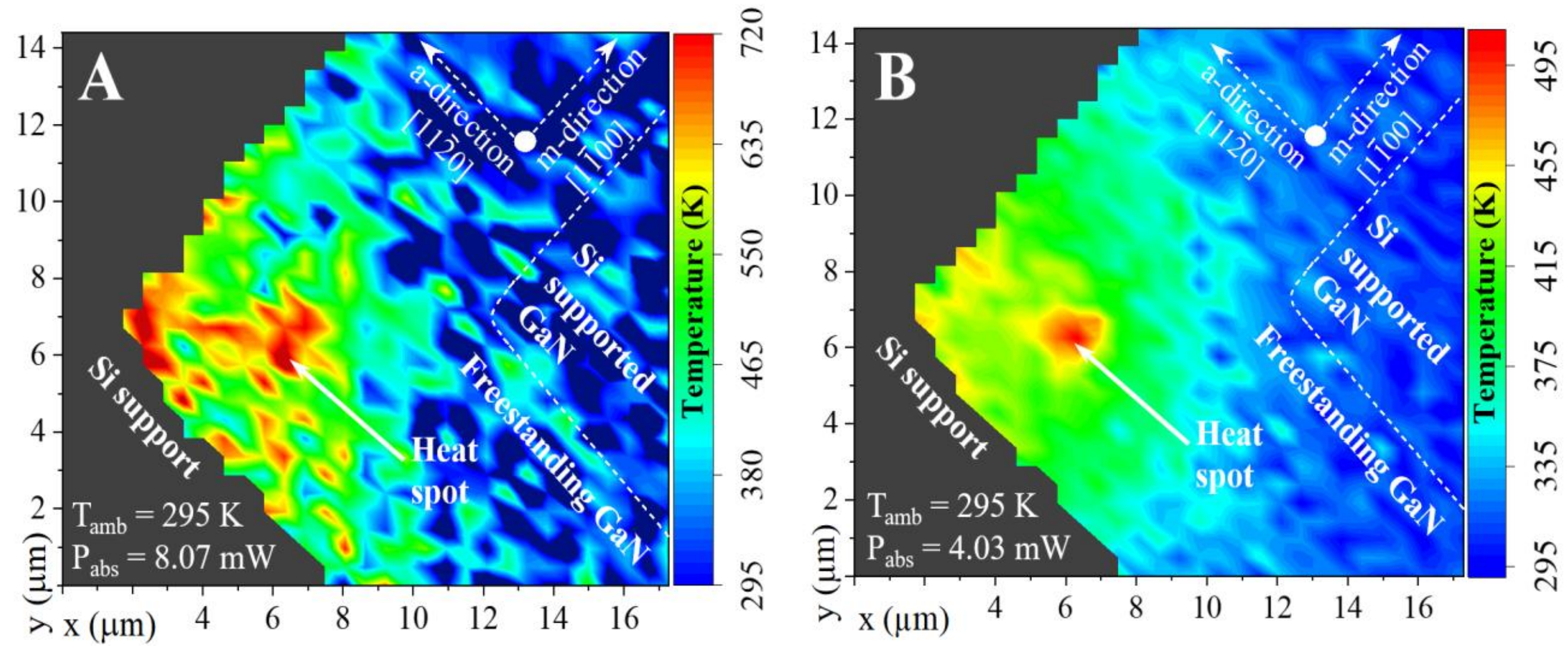

**Fig. S8. Additional 2LRT data measured at the corner (*n* = 2).** (**A**) Temperature map based on the broadening $\Delta\omega$ (FWHM) of the $E_2^{high}$ Raman mode of GaN. This map and the map in Fig. 2B of the main text originate from the same measurement. **(B)** Mid-power ($P_{abs}$ = 4.03 mW) temperature mapscan extracted from the shift $\omega$ of the $E_2^{high}$ Raman mode of GaN. The onset of the edge heating can already be seen.

## Fig. S6

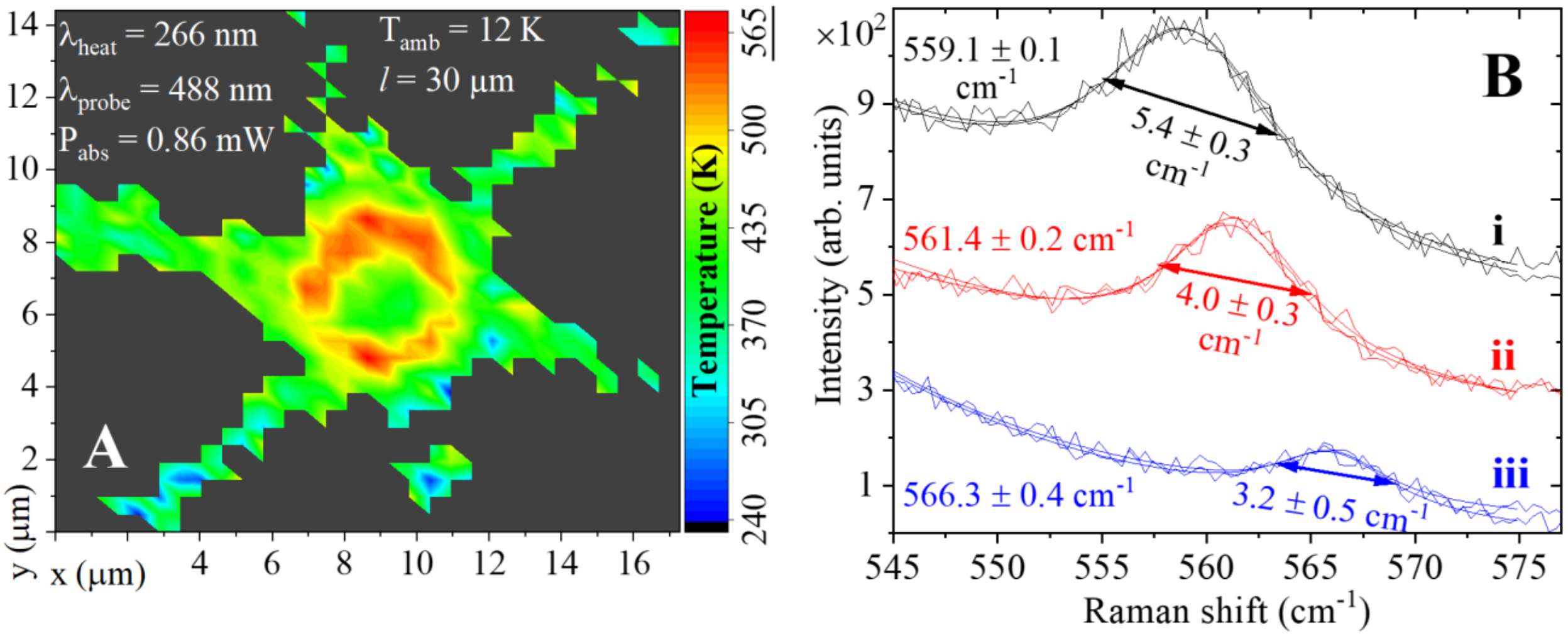

**Fig. S9. Additional 2LRT data – hexagon (*n* = 6).** (**A**) Temperature map based on the broadening $\Delta\omega$ of the $E_2^{high}$ mode of GaN for the freestanding hexagon with $l$ = 30 µm at $T_{amb}$ = 12 K. The data originates from the same measurement shown in Fig. 3A in the main text. In both maps the edge heating is clearly visible. **(B)** Raman spectra showing the optical $E_2^{high}$ mode of GaN at the hot edge (i), in the location of the hot spot (ii) and on the suspending nanobeam (iii) as indicated in Fig. 2F. The edge heating effect is clearly visible in the spectra. The background in the spectra that grows at relative lower wavenumbers stems from the strong Si Raman mode located around 520 $cm^{-1}$.

**Fig. S7**

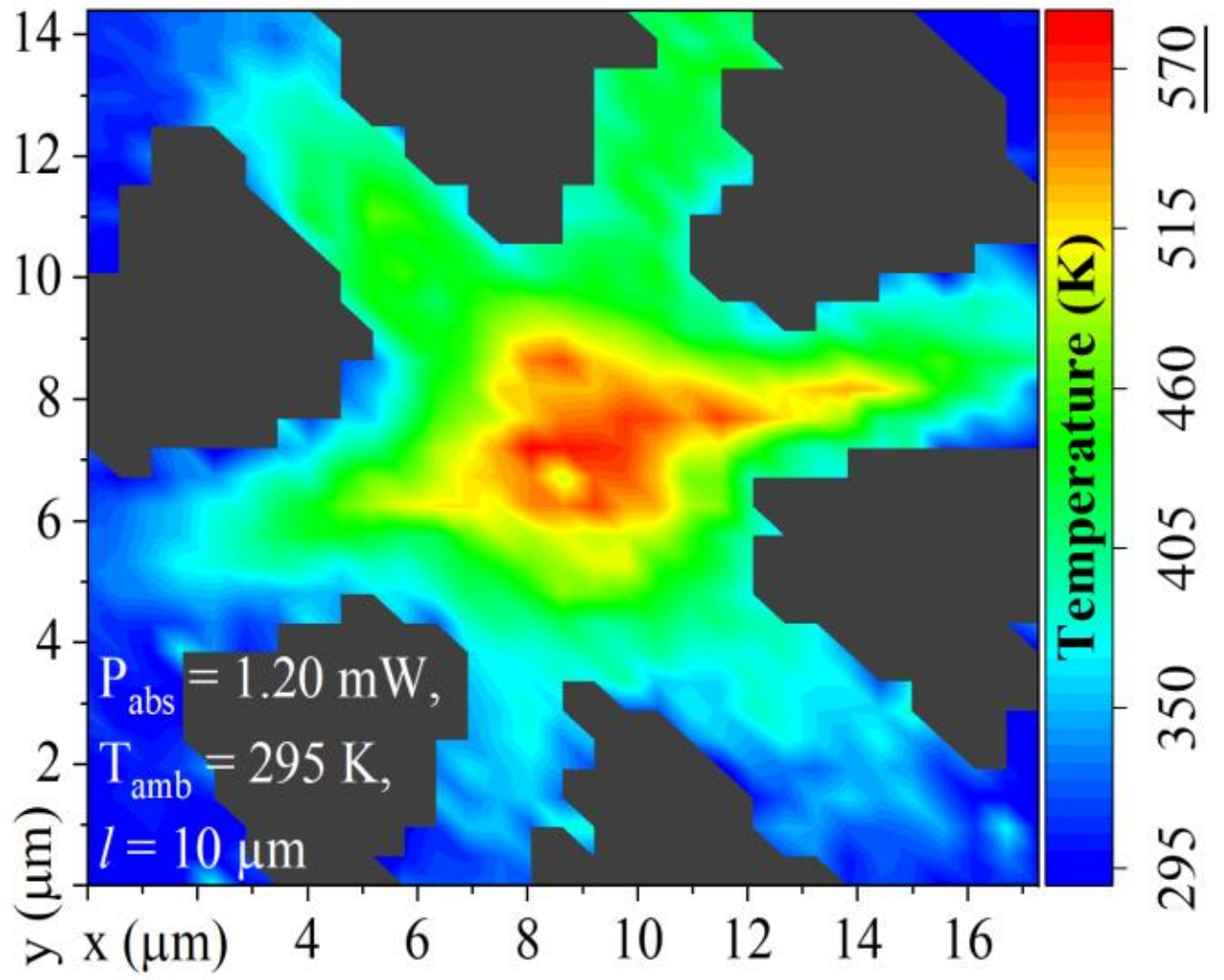

**Fig. S10. Additional 2LRT data – hexagon (*n* = 6) – mid power.** Temperature distribution based on the shift of the $E_2^{high}$ mode of GaN, for a suspended hexagon. $l$ = 10 μm. Due to this relatively short nanobeam length, despite the elevated $P_{abs}$ value, the temperature rise is smaller compared to the $l$ = 15 μm (Fig. 2F) or $l$ = 30 μm (Fig. 3A) hexagon configuration. For the presented set of experimental parameters, no edge heating was observed.

## Fig. S8

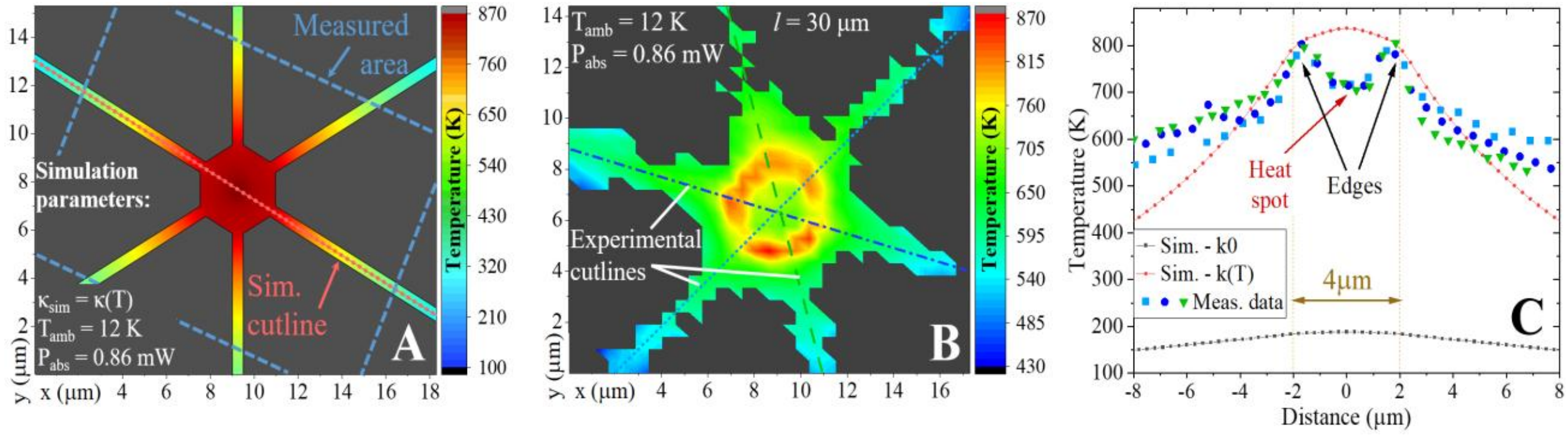

**Fig. S11. Comparison of 2LRT maps to simulations – hexagon (*n* = 6) – Fourier.** (**A**) COMSOL®-based simulation of the suspended hexagon. The simulation parameters, such as the heat and probe beam spot sizes, the $P_{abs}$, and the ambient temperature correspond to the measured experimental parameters. The thermal conductivity is temperature-dependent and has been extracted from Fig. 1A (blue filled triangle) in (7)**,** based on a fit with a logistic function. Details of the simulations can be found in the text of the SM. The red dotted line shows the location where the simulated data was extracted for the comparison with the experiment. **(B)** Temperature mapscan recorded at $T_{amb}$ = 12 K. Note that despite the cooling of the Si substrate, the depicted section reaches temperatures > 430 K. The blue and green lines show the locations where the measured data was taken for comparison with the simulation. The same map is displayed in the main text in Fig. 3A. **(C)** Comparison of the measured data with the simulations. The gray and red dotted lines are COMSOL® simulations based on the heat equation with a constant thermal conductivity $\kappa_0$ = 95 $Wm^{-1}K^{-1}$ (1) and a temperature-dependant $\kappa(T)$, respectively. The light blue, dark blue, and green data points belong to the experimental data. Both simulations fail to qualitatively capture the edge heating. The simulation with a constant thermal conductivity $\kappa_0$ shows a low temperature rise, since it cannot describe the drop in $\kappa$ at elevated temperature. Error bars on the measured data have been omitted for the clarity of the plot.

## Fig. S9

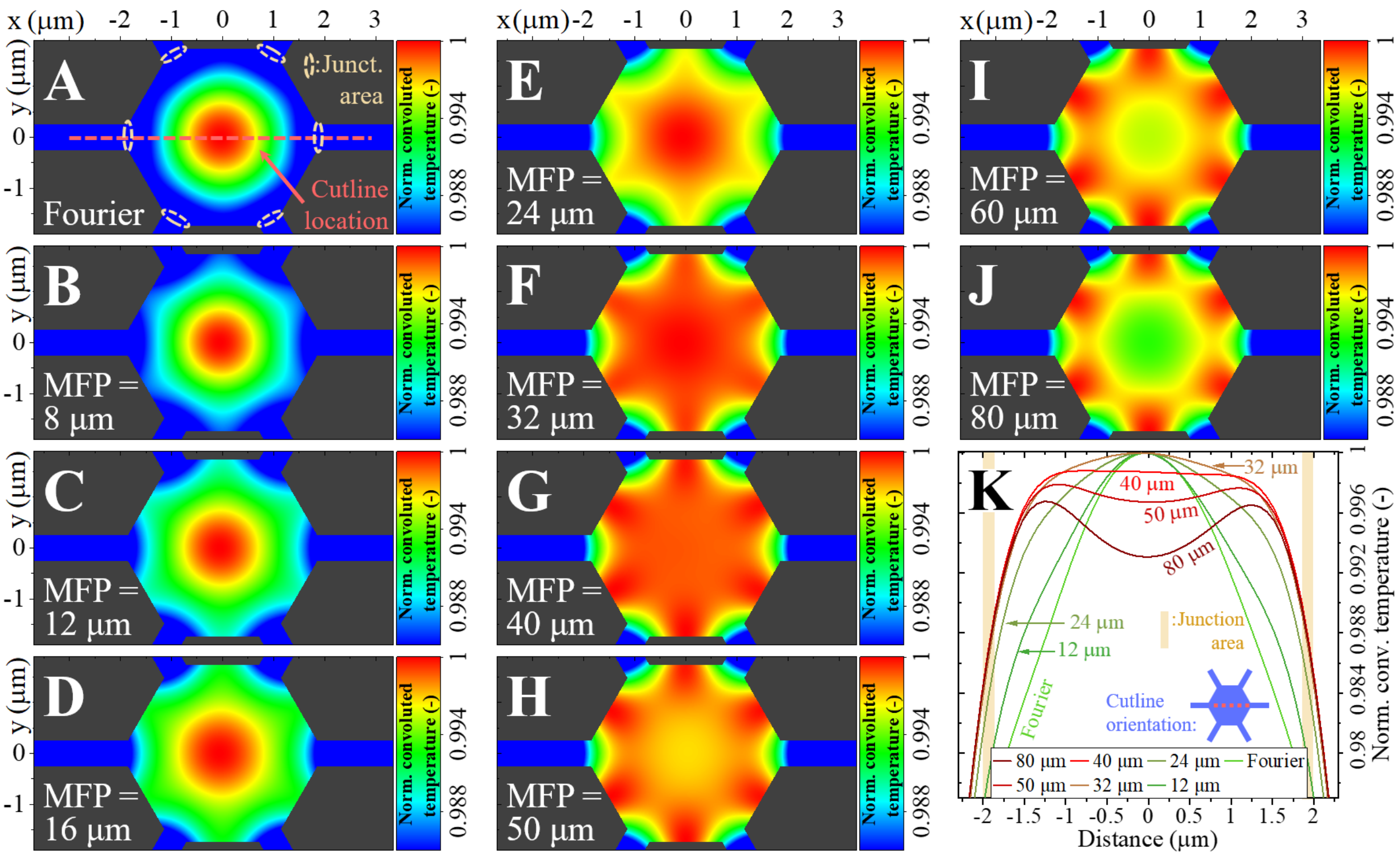

**Fig. S12. 2D-BTE simulations – hexagon ($n$ = 6).** (**A-J**) Close-up of the result of the 2D-BTE simulations of the suspended hexagon ($l$ = 30 µm) convoluted by the probe laser profile function, with MFPs = 8, 12, 16, 24, 32, 40, 50, 60, and 80 µm, as well as the Fourier simulation used to seed our 2D-BTE solver. The probe beam laser has a measured (knife-edge method) FWHM of 1.2 µm, the heat laser a FWHM of 0.5 µm, and the absorbed power is fixed to 0.86 mW. The output has been normalized to allow a qualitative comparison of the various MFP values. In **A**, the location of the cutlines used in **K** is marked by the light-red dashed line. The junction area, which designates the zone where the suspending nanobeams and the suspended hexagon merge, is marked by the dashed beige ovals. (**K**) Comparison of the cutlines through the hexagon in the direction marked in **A** (also depicted in the inset). The vertical beige stripes indicate the location of the junction area. These cutlines are taken in the direction of the nanobeams, contrarily to Fig. 3D, where the cutlines are taken in the perpendicular direction. These images reveal the gradual appearance of the edge heating, suggesting that this phenomenon may appear in a wide variety of materials provided that heat carriers have a sufficiently large MFP.

**Fig. S10**

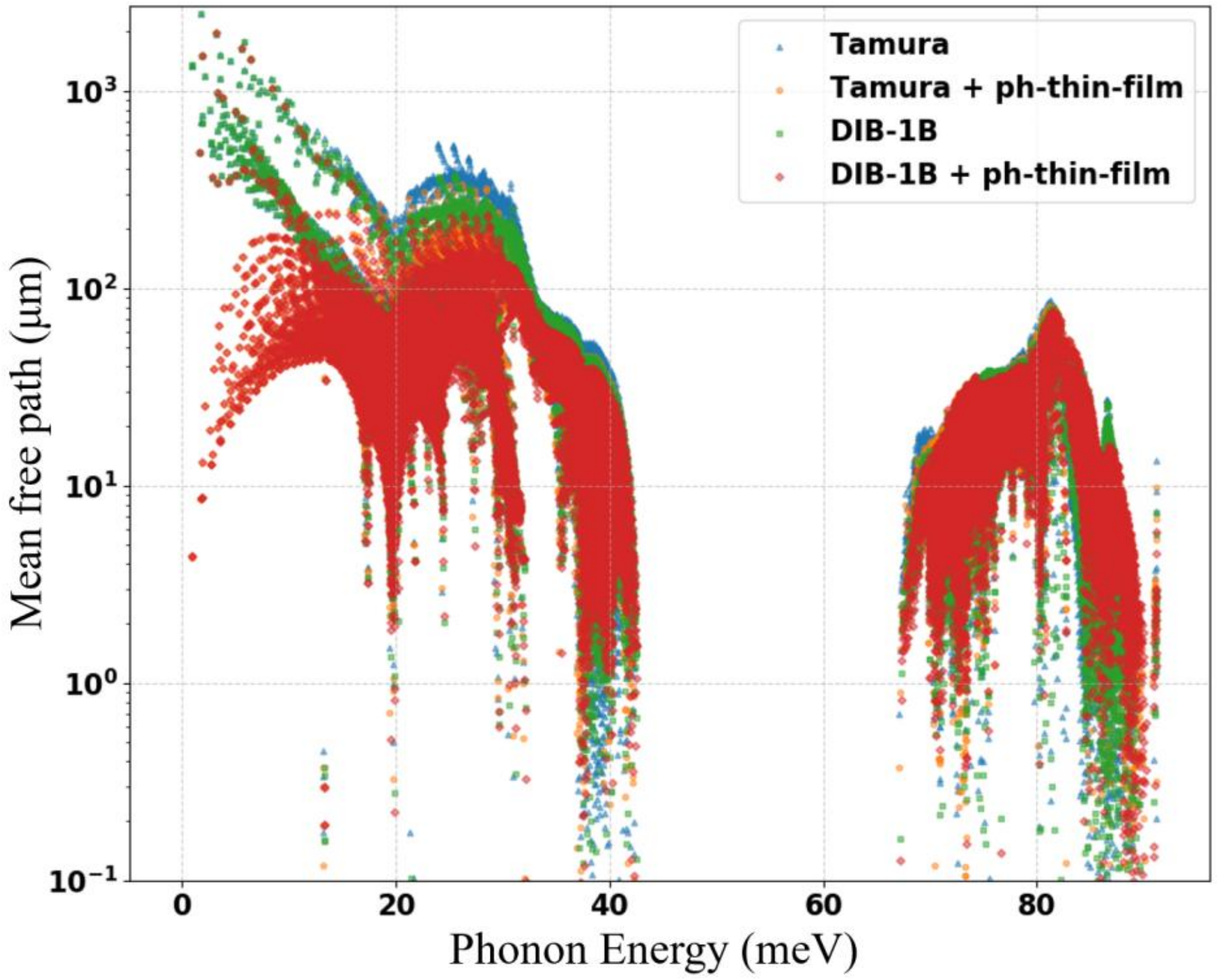

**Fig. S10. Mean free path versus phonon energy calculated for each (s, q) mode, according to Eq. 14.** These distributions were calculated including phonon-isotope interactions modeled with the Tamura and DIB-1B approaches, with and without diffusive phonon-boundary (ph-thin-film) scattering. All calculations included 3- and 4-phonon scattering and were performed for $T$ = 300 K.

## Table S1:

| Mode label | | | | | Mode energy | | | | |
|---|---|---|---|---|---|---|---|---|---|
| $A_1$(LO) [Γ] | → | $E_1$(TO) [L] | + | TA/LA/$E_2^{low}$[L] | 91.6 meV | → | 69.3 meV | + | 22.3 meV |
| $E_1$(LO) [Γ] | → | $E_2^{high}$ [M] | + | TA/LA [M] | 92.4 meV | → | 75.0 meV | + | 17.4 meV |

**Table S1: 3-phonon scattering decay paths of zone center LO modes in wurtzite GaN.** Possible decay paths of zone center LO phonons (17). Only asymmetric decay (resulting phonon with different energies) paths are possible, due to the energy gap from around 42 meV to 77 meV in the phonon-dispersion of GaN (1). By matching the energies of the resulting phonons to the MFPs vs. energy plot shown in Fig. 3F, we notice that these phonons have an MFP of at most 500 nm, which allows us to discard them as candidates to explain the edge heating phenomenon. Only higher order decay paths can generate phonons with larger MFP value (> 1 µm) as sketched in Fig. 3E (green arrows).

## Table S2:

| Interactions | $\kappa$ (Wm$^{-1}$K$^{-1}$) | |
|---|---|---|
| | in-plane | cross-plane |
| Tamura | 248.4 | 252.3 |
| Tamura + ph-thin-film | 136.2 | 87.9 |
| DIB-1B | 219.3 | 222.8 |
| DIB-1B + ph-thin-film | 120.7 | 79.1 |

**Table S2: In-plane ($\kappa_{xx}$, $\kappa_{yy}$) and cross-plane ($\kappa_{zz}$) calculated thermal conductivities.** The sample is oriented in the *xy* plane (perpendicular to the c-axis of wurtzite GaN). Results were obtained by iteratively solving the phonon BTE on a 36 × 36 × 36 wave vector mesh including 3- and 4-phonon scattering, phonon-isotope interactions with the Tamura and DIB-1B models, with and without phonon-boundary (ph-thin-film) scattering. All calculations included 3- and 4-phonon scattering and were performed for $T$ = 300 K.

## References and Notes:

1. M. Elhajhasan, W. Seemann, K. Dudde, D. Vaske, G. Callsen, I. Rousseau, T. F. K. Weatherley, J.-F. Carlin, R. Butté, N. Grandjean, N. H. Protik, G. Romano, Optical and thermal characterization of a group-III nitride semiconductor membrane by microphotoluminescence spectroscopy and Raman thermometry. *Phys. Rev. B* **108**, 235313 (2023).

2. W. Seemann, M. Elhajhasan, J. Themann, K. Dudde, G. Würsch, J. Lierath, G. Callsen, J. Ciers, Å. Haglund, N. H. Protik, G. Romano, R. Butté, J.-F. Carlin, N. Grandjean, Thermal analysis of Ga N -based photonic membranes for optoelectronics. *Phys. Rev. Applied* **25**, 024028 (2026).

3. A. Krost, A. Dadgar, GaN-based optoelectronics on silicon substrates. *Materials Science and Engineering: B* **93**, 77–84 (2002).

4. I. M. Rousseau, III-Nitride Semiconductor Photonic Nanocavities on Silicon. Ph. D. thesis, 2018, École Polytechnique Fédérale de Lausanne (EPFL).

5. G. Romano, OpenBTE: a Solver for ab-initio Phonon Transport in Multidimensional Structures. arXiv arXiv:2106.02764 [Preprint] (2021). https://doi.org/10.48550/arXiv.2106.02764.

6. G. Romano, S. G. Johnson, Inverse design in nanoscale heat transport via interpolating interfacial phonon transmission. *Struct Multidisc Optim* **65**, 297 (2022).

7. Q. Zheng, C. Li, A. Rai, J. H. Leach, D. A. Broido, D. G. Cahill, Thermal conductivity of GaN, GaN 71 , and SiC from 150 K to 850 K. *Phys. Rev. Materials* **3**, 014601 (2019).

8. Ph. Lambin, J. P. Vigneron, Computation of crystal Green's functions in the complex-energy plane with the use of the analytical tetrahedron method. *Phys. Rev. B* **29**, 3430–3437 (1984).

9. N. H. Protik, C. Li, M. Pruneda, D. Broido, P. Ordejón, The elphbolt ab initio solver for the coupled electron-phonon Boltzmann transport equations. *npj Comput Mater* **8**, 28 (2022).

10. W. Li, J. Carrete, N. A. Katcho, N. Mingo, ShengBTE: A solver of the Boltzmann transport equation for phonons. *Computer Physics Communications* **185**, 1747–1758 (2014).

11. Z. Han, X. Yang, W. Li, T. Feng, X. Ruan, FourPhonon: An extension module to ShengBTE for computing four-phonon scattering rates and thermal conductivity. *Computer Physics Communications* **270**, 108179 (2022).

12. S. Tamura, Isotope scattering of dispersive phonons in Ge. *Phys. Rev. B* **27**, 858–866 (1983).

12. P. Virtanen *et al*, SciPy 1.0: fundamental algorithms for scientific computing in Python. *Nat Methods* **17**, 261–272 (2020).

14. T. Beechem, A. Christensen, S. Graham, D. Green, Micro-Raman thermometry in the presence of complex stresses in GaN devices. *Journal of Applied Physics* **103**, 124501 (2008).

15. G. Romano, A. M. Kolpak, Diffusive Phonons in Nongray Nanostructures. *Journal of Heat Transfer* **141**, 012401 (2019).

16. A. Jeżowski, B. A. Danilchenko, M. Boćkowski, I. Grzegory, S. Krukowski, T. Suski, T. Paszkiewicz, Thermal conductivity of GaN crystals in 4.2–300 K range. *Solid State Communications* **128**, 69–73 (2003).

17. D. Y. Song, S. A. Nikishin, M. Holtz, V. Soukhoveev, A. Usikov, V. Dmitriev, Decay of zone-center phonons in GaN with A1, E1, and E2 symmetries. *Journal of Applied Physics* **101**, 053535 (2007).

18. D. Alvarez, JUWELS Cluster and Booster: Exascale Pathfinder with Modular Supercomputing Architecture at Juelich Supercomputing Centre. *JLSRF* **7**, A183 (2021).

**Acknowledgments:**

The authors acknowledge the Gauss Centre for Supercomputing e.V. (www.gauss-centre.eu) for funding this project by providing computing time on the GCS Supercomputer JUWELS (*18*) at the Jülich Supercomputing Centre (JSC). Furthermore, the authors acknowledge J.-F. Carlin for the epitaxy of all III-nitride material used in this study and J. Ordonez-Miranda for valuable discussions.

**Funding:**

N.H.P. and E.T. acknowledge funding from the "Deutsche Forschungsgemeinschaft" (DFG, German Research Foundation) for an "Emmy Noether" research grant (Grant No. 534386252).

G.R. acknowledges funding from the MIT-IBM Watson AI Laboratory (Challenge No. 2415).

G.R. and G.C. acknowledge funding from the MIT Global Seed Funds.

M.E. and G.C. acknowledge funding from the Central Research Development Fund (CRDF) of the University of Bremen for the project "Joint optical and thermal designs for next generation nanophotonics".

The research of M.E., K.D., G.W., J.L. and G.C. was funded by the major research instrumentation program of the DFG (Grant No. 511416444).

The research of I.R. was funded by the Swiss National Science Foundation through Grant No. 200020_162657.

G.C. also acknowledges the MAPEX-CF Grant for Correlated Workflows (Grant No. 40401080) and funding from the MAPEX "Minor Instrumentation Grant" associated to the APF program "Materials on Demand" (MI06/25 and MI07/25).